\documentclass{aa}  

\usepackage{graphicx}
\usepackage{txfonts}
\usepackage{lipsum}
\usepackage{subcaption}         
\usepackage{lscape}             
\usepackage{placeins}           
\usepackage{xspace}
                                
\usepackage[colorlinks=true,linkcolor=red,citecolor=blue]{hyperref}

\newcommand{\sima}{${\sim}$}

\begin{document}
   \title{LoTSS of dual AGN: enhanced detectability and prospects for a systematic search}
   \titlerunning{LoTSS of dual AGN}



   \author{Q. D'Amato\inst{1}\thanks{\email{quirino.damato@inaf.it}}
        \and F. Mannucci\inst{2}
        \and E. De Rubeis\inst{3,4}
        \and I. Prandoni\inst{4,5}
        \and A. De Rosa\inst{1}
        \and M Scialpi\inst{6,7,2}
        \and M. Kunert-Bajraszewska\inst{8}
        \and A. Janiuk\inst{9}
        \and A. Krauze\inst{8}
        \and L. Battistini\inst{10,1}
        \and E. Bertola\inst{2}
        \and C. Bracci\inst{7,2}
        \and A. Chakraborty\inst{11}
        \and M. Ceci\inst{7,2}
        \and C. Cicone\inst{12}
        \and G. Cresci\inst{2}
        \and A. Marconi\inst{7,2}
        \and C. Marconcini\inst{2}
        \and E. Nardini\inst{2}
        \and M. Parvatikar\inst{1}
        \and A. Peca\inst{13,14,15}
        \and K. Rubinur\inst{12}
        \and P. Severgnini\inst{16}
        \and C. Spingola\inst{4}
        \and L. Ulivi\inst{17}
        \and G. Venturi\inst{2}
        \and C. Vignali\inst{18}
        \and M. V. Zanchettin\inst{1}
        }

   \institute{INAF - Istituto di Astrofisica e Planetologia Spaziali, Via del Fosso del Cavaliere, 00133 Roma, Italy
   \and INAF - Osservatorio Astrofisico di Arcetri, Largo E. Fermi 5, 50125 Firenze, Italy
   \and University of Hamburg, Gojenbergsweg 112, 21029 Hamburg, Germany
   \and INAF - Istituto di Radioastronomia, Via Piero Gobetti 101, 40129, Bologna, Italy
   \and Department of Astronomy, University of Cape Town, 7701 Rondebosch, Cape Town, South Africa
   \and University of Trento, Via Sommarive 14, Trento, 38123, Italy
   \and Dipartimento di Fisica e Astronomia, Università di Firenze, Via G. Sansone 1, Sesto Fiorentino, 50019, Italy
   \and Institute of Astronomy, Faculty of Physics, Astronomy and Informatics, NCU, Grudziądzka 5/7, 87-100, Toruń, Poland
   \and Center for Theoretical Physics, Polish Academy of Sciences, Al. Lotników 32/46, 02-668, Warsaw, Poland
   \and Dipartimento di Matematica e Fisica, Università degli Studi Roma Tre, via della Vasca Navale 84, 00146 Roma, Italy
   \and Indian Institute of Astrophysics, 2nd Block, Koramangala,  Bangalore 560 034, India
   \and Institute of Theoretical Astrophysics, University of Oslo, PO Box 1029, Blindern, Oslo, 0315, Norway
   \and Universidad Diego Portales, Facultad de Ingenier\'ia y Ciencias, Instituto de Estudios Astrof\'isicos, Av. Ej\'ercito Libertador 441, Santiago, Chile
   \and Eureka Scientific, 2452 Delmer Street, Suite 100, Oakland, CA 94602-3017, USA
   \and Department of Physics, Yale University, P.O. Box 208120, New Haven, CT 06520, USA
   \and INAF - Osservatorio Astronomico di Brera, via Brera 28, Milano, 20141, Italy
   \and Centro de Astrobiología, CSIC--INTA, Cra. de Ajalvir Km. 4, Torrejón de Ardoz, Madrid, 28850, Spain
   \and Dipartimento di Fisica e Astronomia, Università degli Studi di Bologna, Via Gobetti 93/2, Bologna, 40129, Italy
   }

   \date{Received September 30, 20XX}

 
\abstract
{Dual active galactic nuclei (DAGN) -- systems of two simultaneously accreting supermassive black holes at kpc-scale separations -- are critical tracers of galaxy mergers and the progenitors of gravitational-wave sources targeted by pulsar timing arrays and the upcoming Laser Interferometer Space Antenna. Despite their importance, their radio properties remain largely unconstrained, particularly at high redshift, due to the limited combination of sensitivity and sky coverage of existing high-angular-resolution surveys.}
{In this pilot project, we primarily aim at demonstrating the detectability of high-redshift DAGN at 144 MHz performing sub-arcsec resolution imaging of the LOFAR Two-metre Sky Survey (LoTSS) data, and characterize the low-frequency radio properties of individual systems.}
{We cross-matched a sample of 92 spectroscopically confirmed DAGN at $z > 0.3$ and projected separation $r_p < 30$ kpc with the LoTSS catalogs of data release (DR) 2 (resolution $\sim$6\arcsec), and reprocessed the international LOFAR telescope (ILT) LoTSS data for three sources spanning $0.59 < z < 2.39$, achieving $\sim$0.3\arcsec\ resolution at 144 MHz. We performed multi-frequency analysis and brightness temperature diagnostics to characterize the AGN nature and radio properties of each system.}
{We successfully produced high-quality ILT images for the three DAGN in the pilot sample. In one source at $z = 1.49$, we detect both AGN components at $>5\sigma$, making it one of the highest-$z$ and faintest DAGN detected at radio frequencies. Brightness temperature measurements confirm that ILT 144 MHz observations can efficiently distinguish AGN-powered emission from star formation even in sub-mJy sources. In the other two sources we detect only one AGN. In one of them, the new ILT imaging constrains the low-frequency radio spectrum of the compact steep spectrum source, indicating that its complex morphology is shaped by jet-ISM interaction rather than by the companion galaxy.
In addition, for the full 92 DAGN sample, we found a detection rate of $\sim$24\% in both LoTSS DR2 and DR3, significantly larger than that of single AGN matched in selection ($\sim$9--10\%), suggesting an enhanced detectability of DAGN in the LoTSS.
}
{Low-frequency, sub-arcsec resolution radio observations constitute a highly efficient and dust-unbiased pathway to detect, confirm, and characterize DAGN across cosmic time. The pilot sample presented here establishes the foundation for a systematic DAGN search in the upcoming all-sky $\sim$0.3\arcsec\ resolution iLoTSS survey and provides an early benchmark for future SKA observations.}


   \maketitle

\nolinenumbers
\section{Introduction}
\label{sec:intro}
Hydrodynamical simulations show that major galaxy mergers can host pairs of two supermassive black holes (SMBHs) that will eventually merge with long timescales ($\sim$1 Gyr, \citealt{tremmel_2017}). Massive, merger-driven gas inflows are also expected to efficiently fuel both SMBHs, leading to enhanced accretion activity and the formation of dual active galactic nuclei (DAGN; e.g., \citealt{volonteri_2003,colpi_2014,volonteri_2022,chen_2023_sim,shen_2023}). DAGN are systems in which two simultaneously accreting SMBHs have separations between hundreds of pc and a few tens of kpc (i.e., larger than their radii of gravitational influence) evolving within the overall potential of their merging host galaxies \citep{merritt_2013,volonteri_2022}. These are critical systems because they are the progenitors of the inspiralling or merging SMBHs at the origin of low-frequency (nHz-$\mu$Hz) gravitational waves (GWs) that are the primary targets of Pulsar Timing Array (PTA) projects (e.g., \citealt{agazie_2023,EPTA_collab_2023}) and of the upcoming Laser Interferometer Space Antenna (LISA, \citealt{Amaro-Seoane_2023,colpi_2024}) observatory. Building statistical samples of DAGN and studying their properties is pivotal to test galaxy evolution scenarios (e.g., \citealt{volonteri_2022,shen_2023,dimatteo_2023}). However, observational identification of these systems is a difficult task, especially at high redshift ($z\geq0.3$), due to their paucity and because this effort requires covering large areas of sky with sub-arcsec resolution \citep{derosa_2019, severgnini_2026}. Modern selection methods usually rely on relatively bright optically-selected quasars identified in \textit{Gaia} and Euclid  surveys (\citealt{shen_2019,mannucci_2022,ulivi_2025, gaia_collab_2024}), which offer precise astrometry, high angular resolution, and large sky coverage; however, the optical emission can be associated to DAGN, gravitationally lensed systems, or projected alignment between an AGN and a foreground star, and their efficiency in finding genuine DAGN systems is 10--50\%, depending on their separation, magnitude, and Galactic latitude \citep{scialpi_2025}. X-ray emission has also been used to select DAGN, but due to low photon-counting statistics and poor spatial resolution on wide areas, the detection and characterization of a substantial population of DAGN is practically feasible only in the nearby Universe \citep{derosa_2023,battistini_2026}. In addition, these optical and X-ray selection techniques are biased towards dust-free and mildly unobscured objects \citep{blecha_2018}, preventing us from obtaining a complete census of the DAGN population.

Radio emission can, in principle, serve as a powerful tool for identifying DAGN candidates: it is unaffected by dust obscuration, can reach sub-arcsec resolution through interferometry (down to milli-arcsec, mas, scales with very long baseline interferometry, VLBI), and is not contaminated by photoionised gas emission that can mimic a second AGN in optical observations (e.g., \citealt{fu_2012}). Modern radio interferometers can also reach $\sim\mu$Jy sensitivities and survey large sky areas with relatively modest time investment compared to other wavelengths. However, current cm-wavelength surveys with sub-arcsec resolution are limited by either shallow sensitivity or small area coverage \citep{smolcic_2017, damato_2022}, and radio emission in the sub-mJy regime can originate from either AGN or star formation (SF) activity. VLBI observations (mas resolution) are currently required to confirm DAGN, including binary systems, through the detection of multiple compact radio cores at kpc- or pc-scale separations \citep{rodriguez_2006,kharb_2017,damato_2026}. However, VLBI narrow (typically arcsec-scale) field of view makes them inefficient for blind DAGN searches. As a result, only a handful of DAGN have been confirmed and studied at radio wavelengths to date \citep[e.g.,][]{fu_2011,deane2014nature,rubinur2019, spingola2019,chen_2023_vodka,glikman_2023,xu_2024,schwartzman_2024}, and the radio properties of compact, high-$z$ DAGN remain poorly known.

The cleanest way to unambiguously identify AGN-dominated radio emission at a single frequency, separating it from SF, is via the equivalent brightness temperature ($T_\mathrm{B}$), defined as the temperature of a black body that would produce the observed surface brightness (flux density per solid angle) at a given frequency. If the measured $T_\mathrm{B}$ significantly exceeds the value expected from SF processes,  \citep[$T_\mathrm{B,SF}$,][]{condon_1992}, the source is reliably AGN-dominated. $T_\mathrm{B,SF}$ ranges from $\sim10^6$ K at $\sim$100 MHz to $\sim10^5$ K at $\sim$1 GHz \citep{condon_1992, morabito_2022}, making sub-arcsec observations at a few hundred MHz especially effective for $T_\mathrm{B}$-based AGN selection. For example, a point-like $z=0$ source with $1$ mJy at 100 MHz, already reaches $T_\mathrm{B}\sim10^6$ K if observed at $\sim$0.3\arcsec\ resolution. At 1 GHz, the same source would have a flux density of $\sim0.25$ mJy, assuming a synchrotron spectral slope $\alpha=-0.7$ ($S_\nu \propto \nu^\alpha$), and would require a resolution of $\sim$50 mas  to reach the $T_\mathrm{B}\sim10^5$ K threshold.
This method has been successfully applied to high-resolution ($\sim$0.3\arcsec) LOw-Frequency ARray (LOFAR, \citealt{van_haarmel_2013}) observations at 144 MHz, demonstrating its effectiveness in selecting AGN and uncovering a new low-frequency AGN population \citep{morabito_2022}.

The LOFAR Two-metre Sky Survey (LoTSS, \citealt{shimwell_2017}) is an ongoing 144 MHz LOFAR continuum observational program aiming at covering the whole northern sky down to a 5$\sigma$ flux limit of $\sim$500 $\mu$Jy/beam. Although only Dutch stations are included in the publicly released data products (providing  6\arcsec\ resolution images), 
LoTSS observations also include the international LOFAR telescope (ILT) stations, with baselines of thousands of km. Their addition enables the production of images with significantly better resolution (down to 0.3$''$).  A resolution of 0.3$''$ implies the possibility to disentangle two (or more) AGN components down to a projected separation of $\sim$2\,kpc at $z\gtrsim$0.5, with a maximum of $\sim$2.5 kpc at $z\sim1.5$. Hence, this survey is perfectly suited to perform a systematic search and study of the DAGN population at radio frequencies.

To characterize the DAGN population at low radio frequencies and evaluate the effectiveness of this approach in identifying new systems, we initiated a pilot project aimed at reprocessing the LoTSS data, including the ILT, to reach 0.3$''$ resolution. For clarity, throughout this paper we refer to the 6\arcsec\ resolution images as ``LOFAR data'' and to the 0.3\arcsec\ resolution images as ``ILT data'', while we use the term LoTSS to generally refer to the survey. As a proof of technical feasibility and to illustrate the potential outcomes, here we present high-resolution imaging and analysis for an initial subsample of three sources, spanning a wide range in redshift, flux density, and distance from the pointing centers of LoTSS fields.

In Sec. \ref{sec:selection} we describe the selection of the targets and present their properties. The LoTSS data reduction and high-resolution imaging of the three-objects subsample is described in Sec. \ref{sec:data_redux}, while is Sec. \ref{sec:results} we present the measurements and quantities derived from the ILT imaging. The analysis and discussion of each target is presented in Sec. \ref{sec:ind_sources}. In Sec. \ref{sec:fut_per} we statistically assess the excess of DAGN detection rate with respect to single AGN in the LoTSS, and present the future perspectives and potential outcomes of the pilot projects. Finally, in Sec. \ref{sec:concl} we draw the conclusions of this work. Throughout the work we adopt a concordance $\Lambda$CDM cosmology with $\mathrm{H_0} = 70~\mathrm{km~s^{-1}~Mpc^{-1}}$, $\Omega_{\mathrm{M}} = 0.3$, and $\Omega_{\mathrm{\Lambda}} = 0.7$, in agreement with the \textit{Planck 2018} results \citep{PLANCK_2018}.

\section{Target selection and description}
\label{sec:selection}
Since this pilot project aims at demonstrating the detectability of high-$z$ DAGN in ILT images and studying their low-frequency properties, we started the selection from a sample of known DAGN. Namely, we considered all the 92 DAGN at $z>0.3$ with projected physical separation $r_\mathrm{p}< 30$ kpc that were spectroscopically confirmed to the best of our knowledge as of January 2025 (see Table \ref{tab:full_sample}). These redshift and separation ranges are chosen to consider systems that are or will be involved in an interaction according to theoretical predictions \citep{comeford_2013, capelo_2015, silverman_2020, volonteri_2022}. Among these, 25 are in the footprint of LoTSS data release 2 (DR2, \citealt{shimwell_2022}) (covering 27\% of the northern sky for a total of $\sim5600~ \mathrm{deg}^2$ at 6$''$ resolution), and 6 of them are detected and present in the LoTSS DR2 catalog (5$\sigma$ flux limit of $\sim500~\mu$Jy/beam). The 92 DAGN in the sample span an angular separation 1.7\arcsec\ $\lesssim \theta_\mathrm{_{p}} \lesssim$ 3.8\arcsec, thus for all of them the two AGN can not be disentangled in the original LOFAR 6\arcsec\ resolution images of the LoTSS.

To demonstrate technical feasibility, in this work we performed data reduction and sub-arcsec resolution imaging (see Sec. \ref{sec:data_redux}) of an initial subsample of three objects. Our ability to perform high-resolution imaging critically depends on source brightness, distance from the pointing center, and separation from the in-field delay calibrator. We therefore selected targets spanning wide ranges in total flux density (two orders of magnitude), pointing-center distance ($0.5-1.2$ deg), and delay calibrator distance ($0-0.5$ deg), that is close to the limits allowed by the inclusion of international baselines \citep{sweijen_2022}. The three sources also span a wide range of redshift ($0.6 \lesssim z \lesssim2.4)$ and similar angular separation between the nuclei ($\sim 1.5 - 2.3$''). The properties of the sources are listed in Table \ref{tab:src_prop}.

J0930+4614 at $z=2.394$ was first selected as a lens candidate on the basis of morphological and color criteria in the Sloan Digital Sky Survey (SDSS, \citealt{gunn_2006, oguri_2006}), and confirmed as a DAGN through spectroscopic observations in the SDSS-III Baryon Oscillation Spectroscopic Survey
(BOSS, \citealt{more_2016}), which revealed very different optical spectra for the two objects. The primary optical AGN (AGN A) is detected by the Very Large Array (VLA) at 1.4, 4.9, 8.5 and 22.4\,GHz, with images available at the NRAO VLA archive survey (NVAS, \citealt{crossley_2007}). It is also detected by the Very Long Baseline Array (VLBA) at 4.3 GHz and 7.6 GHz,  with images available at the Astrogeo VLBI FITS image database\footnote{\url{https://astrogeo.org/vlbi_images/}}. The secondary optical AGN (AGN B, at an angular separation of 1.5\arcsec) is not detected in any of these radio images. 

J1120+6711 at $z=1.490$ has been selected as a lens candidate on the basis of color criteria in the SDSS, and followed up with the Arizona Infrared Imager and Echelle Spectrograph (ARIES) at the Multiple Mirror Telescope. The markedly different optical spectra led to its DAGN classification \citep{pindor_2006}. No other radio observations of this DAGN are available.

J1643+3156 was discovered by \cite{brotherton_1999}. The source was identified as a dual system at $z=0.589$ by low resolution imaging spectrometer (LRIS) observations with the Keck telescope, and subsequently observed with the  Karl G. Jansky Very Large Array (JVLA) at 8.2 GHz with \sima0.7$''$ resolution, allowing a clear detection of the primary AGN (AGN A) and a tentative detection of the secondary AGN (AGN B), at angular separation of 2.3$''$. On the basis of its compactness (\sima7 kpc) and global (spatially-unresolved) steep spectral index ($\alpha\sim-0.7$), \cite{brotherton_1999} classified AGN A as a compact steep spectrum (CSS; \citealt{odea_1998}) source, a class of small-scale ($<$20 kpc) young radio sources. The source also has optical and X-ray properties consistent with the CSS population. J1643+3156 has been subsequently re-observed at 1.6\,GHz and 5\,GHz with the enhanced Multi Element Remotely Linked Interferometer Network (eMERLIN) by \citet{kunert-bajraszewska_2011} and at 10\,GHz with the JVLA \citep{Maithil_2020}. In particular, on the basis of high-resolution (beam $\lesssim 0.2''$) eMERLIN and Hubble space telescope (HST) observations, \cite{kunert-bajraszewska_2011} have investigated scenarios in which the interaction with the companion galaxy possibly plays a role in modulating the AGN activity (see Sec. \ref{sec:J1643}).

\begin{table*}
\caption{\label{tab:src_prop} Summary of the targets and ILT imaging at 0.3$''$ resolution.
}
\centering
\resizebox{\hsize}{!}{
\begin{tabular}{cccccc|cccc}
\hline\hline

ID & RA(J2000) & DEC(J2000) & $z$ & $\theta_\mathrm{_{p}}$ ($''$) & $r_\mathrm{_{p}}$ (kpc) & Briggs & Beam & Peak flux (mJy) & $\sigma_\mathrm{rms}$ ($\mu$Jy/beam) \\
(1) & (2) & (3) & (4) & (5) & (6) & (7) & (8) & (9) & (10)\\
\hline

\object{$\mathrm{J0930+4614}$} & $09^h 30^m 21^s.16$ & $+46^{\circ} 14' 22''.80$ & 2.394 & 1.5 & 12.2 & 0 & $0.54'' \times 0.37''$ & 23.7 & 62 \\
\object{$\mathrm{J1120+6711}$} & $11^h 20^m 12^s.11$ & $+67^{\circ} 11' 16''.00$ & 1.490 & 1.7 & 14.4 & 0 & $0.53'' \times 0.37''$ & 1.4 & 55 \\
\object{$\mathrm{J1643+3156}$} & $16^h 43^m 11^s.33$ & $+31^{\circ} 56' 19''.00$ & 0.586 & 2.3 & 15.2 & $-1$ & $0.27'' \times 0.15''$  & 39.0 & 99 \\

\hline
\end{tabular}
}
\tablefoot{
(1) Source ID.
(2) Source right ascension (RA) and (3) declination (DEC).
(4) Redshift.
(5) Projected angular separation.
(6) Projected physical separation.
(7) Briggs parameter.
(8) Synthesized beam size.
(9) Peak flux density.
(10) Root-mean-square noise.
}
\end{table*}
 
\section{Data calibration and imaging}
\label{sec:data_redux}
The targets are located in different pointings observed as part of the LoTSS with the high band antenna (HBA) and including ILT stations, covering the frequency range $\sim$120--168 MHz up to a maximum baseline of $\sim$2000 km. Each LoTSS pointing observation is 8 h long. The analyzed pointings in this work are P141+47, P171+67, and P230+33 for J0930+4614, J1120+6711, and J1643+3156, respectively. Pointing data were retrieved from the LOFAR Long-Term Archive (LTA; projects IDs: LC9\_019 and LT10\_010). We also downloaded from the LTA the data of primary calibrators needed for the initial calibration phase, and observed for $\sim10$ min within $\sim24$ h from the target observation. Primary calibrators are 3C196, 3C48 and 3C295 for J0930+4614, J1120+6711, and J1643+3156, respectively. 

\subsection{Initial calibration}
We perform the first step of calibration by running the LOFAR initial calibration (LINC) pipeline \citep{van_weeren_2016, williams_2016, de_gasperin_2019} for both the primary calibrator and pointing observations. For the primary calibrator, LINC corrects for polarization alignment, inter-station clock delays, and bandpass effects for both Dutch and international stations, using internal models of the calibrators \citep{scaife_2012}. Sub-bands or antennas showing anomalous behavior in the bandpass or amplitude--flag diagnostic plots are identified and excluded before proceeding to the next step. The primary calibrator solutions are then parsed to the target pointings, and LINC computes the correction for direction-independent effects by calibrating the phases against a sky model of the pointing region, which is obtained from the TIFR GMRT Sky Survey (TGSS, \citealt{intema_2017}). At the upper end of the HBA band ($\gtrsim$165\,MHz), strong radio-frequency interference (RFI) typically results in full flagging, so sub-bands above this threshold are excluded from all subsequent steps to avoid pipeline failures. A quick-look image of the pointing field is produced and inspected to confirm the absence of  residual artifacts. 

\subsection{Calibration of the international stations}
We followed the general calibration strategy described by \cite{morabito_2022_calib}, suited to calibrate HBA data using the full ILT array and image individual sources at sub-arcsec resolution at 144 MHz. This pipeline initially applies the Dutch-array solutions to all baselines, including the international stations, and checks for contamination of known bright sources in the sky, which in our case are far enough ($>$ 30 deg) not to significantly contaminate our pointings. It then splits the data into sub-bands of 1.95 MHz each, which are phase-shifted onto an in-pointing delay calibrator (ideally within $\sim0.7$ deg distance from the target) that is used to solve for dispersive delays. We searched for suitable calibrators in the long baseline calibrator survey (LBCS; \citealt{jackson_2016,jackson_2022}), finding that J1643+3156 is a calibrator itself, due to its high flux density. For the other two objects, we have chosen as calibrators two sources named L334388 and L440528 in the LBCS, which are distant $\sim$0.5 deg and $\sim$ 0.3 deg from 0930+4614 and J1120+6711, respectively.
The solutions are derived from the delay calibrators using iterative self-calibration with \texttt{facetselfcal} \citep{van_weeren_2021}, starting from a source model obtained from the Very Large Array Sky Survey (VLASS; \citealt{lacy_2020}). \texttt{facetselfcal} is based on \texttt{Default Preprocessing Pipeline} (DP3, \citealt{van_diepen_2018,dijkema_2023}) and \texttt{WSClean} \citep{offringa_2014,offringa_2017}, and calculates both phase and amplitude solutions at each step, updating them in the model of the next self-calibration cycle. We followed a procedure similar to that described by \cite{de_rubeis_2025}. Specifically, during the self-calibration cycles, we solved for \texttt{scalarphasediff} (time interval of 128 s for all calibrators, frequency smoothness of 4 MHz for J0930+4614 and J1643+3156, and 10 MHz for J1120+6711), \texttt{scalarphase} (time interval of 32 s for all calibrators, frequency smoothness of 1 MHz for J0930+4614 and J1643+3156, and 10 MHz for J1120+6711) and \texttt{scalarcomplexgain} (time interval of 53 min and frequency smoothness equal to 1 solution every 10 channels for J0930+4614 and J1643+3156, and time interval of 1\,h and frequency smoothness of 10 MHz for J1120+6711). We excluded all baselines shorter than 40k$\lambda$ equivalent to an angular scale of $\sim$5$''$ at 144\,MHz, in order to avoid potential incompleteness in the sky model, and constrained solutions of \texttt{scalarphasediff} to be the same for the Dutch stations. We phased up LOFAR’s core stations to create a large virtual station with a narrow field of view, thereby reducing interference from unrelated nearby radio sources, particularly on short baselines. We also averaged data down to 32\,s in time and 488\,kHz in frequency, and used a Briggs weighting scheme \citep{briggs_1995} with robust parameter equal to $-1$ in the imaging for each cycle. Default-recipe \texttt{AOFlagger} \citep{offringa_2012} is applied before self-calibration to mitigate residual RFI. The self-calibration is iterated until no significant improvement in image quality is observed.

\subsection{Self-calibration and imaging of the targets}
\begin{figure*}[h!]
        \centering
\resizebox{\hsize}{!}
{\includegraphics[]{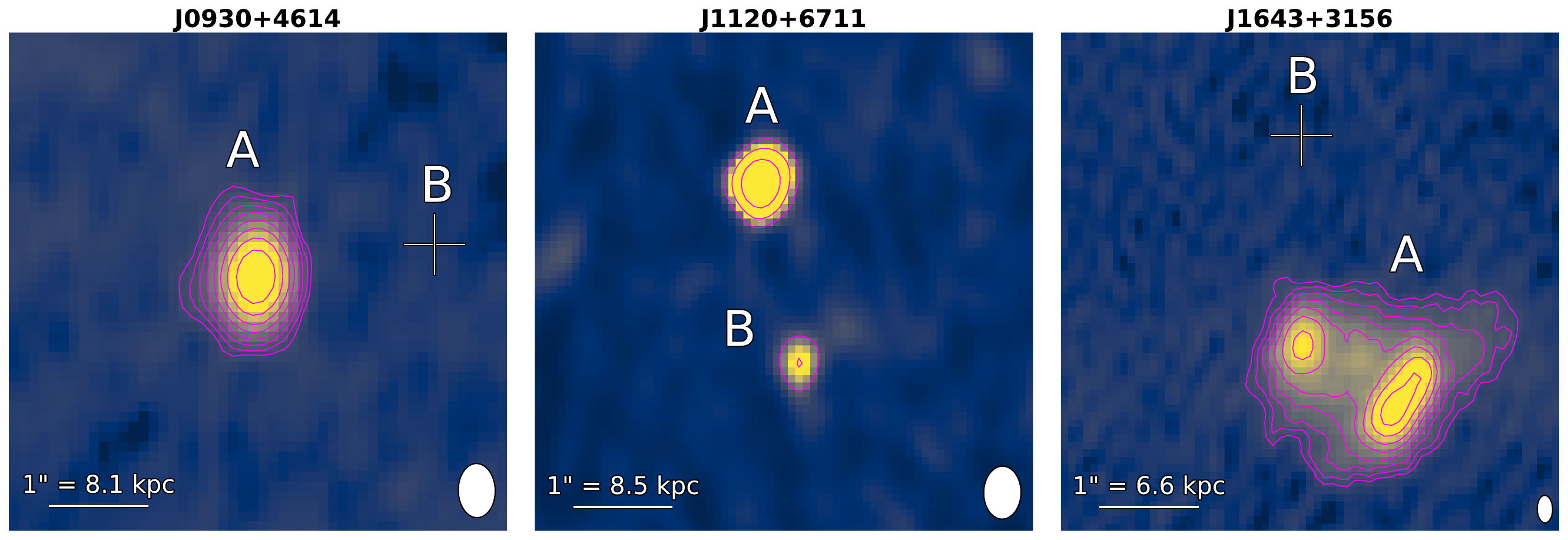}}

        \caption{ILT 144 MHz maps of the targets. Emission contours are overlaid in magenta, starting at the ${\sim}5 \sigma$ level and increasing with a $\sqrt{3}$ geometric progression. The solid white line in the bottom-left corner indicates the angular and physical scale. The white ellipse in the bottom-right corner indicates the restoring beam. In the image of J1120+6711 both AGN are detected. In the images of J0930+4614 and J1643+3156 only one AGN is detected, and the white crosses mark the positions of the undetected companions. Primary and secondary AGN are labeled as A and B, respectively.}
        
        \label{fig:lofar_images}
\end{figure*}

We performed this last step only for J0930+4614 and J1120+6711, since J1643+3156 is a calibrator itself and we used the image produced by the previous step, after 7 cycles of self-calibrations. For both sources, we used \texttt{facetselfcal} to pre-apply the delay calibrator solutions and set a minimum baseline of 20k$\lambda$, while \texttt{WSClean} is used for the imaging. For J1120+6711 the subsequent self-calibration was not possible due to the faintness of the source, and cycle 0 image is used. For J0930+4614, we performed the self-calibration by solving at each cycle for \texttt{scalarphase} (time interval of 32\,s, frequency smoothness of 10\,MHz) and \texttt{scalarcomplexgain} (time interval of 1.1\,h, frequency smoothness of 5\,MHz); the best image is obtained at the 8th cycle.
The final images of J0930+4614 and J1120+6711 were produced using a Briggs robust parameter equal to 0, to achieve the best balance between sensitivity and resolution, while we used the robust $= -1$ image produced by the previous step for J1643+3156 to achieve a smaller restoring beam and resolve the substructures of its complex morphology. The restoring beams and basic properties of the images are reported in Table \ref{tab:src_prop}. 
We established a robust procedure that produced high-quality 0.3$''$ images for all targets, confirming technical feasibility (Fig. \ref{fig:lofar_images}). In one case (J1120+6711), we identified emission from both AGN in the system. This source is the faintest ($\sim$3 mJy) and at the highest redshift ($z\sim1.5$) in the sample, moderately far from the pointing center ($\gtrsim$0.5 deg) and delay calibrator ($\gtrsim$0.3 deg), and with an angular separation of only $\sim$1.7$''$, demonstrating the strong potential of this technique for identifying DAGN in the LoTSS.

\section{Results}
\label{sec:results}
We measured the total flux density $S_{144~\mathrm{MHz}}$ of the detected sources in the ILT 144 MHz images by performing a Gaussian fitting of their emission with the Cube Analysis and Rendering Tool for Astronomy (CARTA; \citealt{comrie_2021}). For the undetected ones, we report a $5\sigma$ upper limit from the image root-mean-square $\sigma_\mathrm{rms}$.
Throughout the text, flux density ($S_\nu$) uncertainty is computed as:
\begin{equation}\label{eq:flux_err}
\sigma_{S_\nu}= \sqrt{(A_\mathrm{int}~\sigma_\mathrm{rms})^2+(S_\nu ~ \sigma_\mathrm{cal})^2},
\end{equation}
where $A_\mathrm{int}$ is the integration area in beam units, and $\sigma_\mathrm{cal}$ is the calibration error conservatively assumed to be 20\% for the 144 MHz images and 10\% for the GHz-range archival images when they are used \citep{perley_2013,sweijen_2022}. 
J0930+4614 and J1120+6711 are well fitted by a single Gaussian component, while for the extended irregular emission of J1643+3156 we found that a 6-Gaussian components fitting provided the best results (the total flux density is 1$\sigma$ in agreement with that measured manually within the $5\sigma$ contours). The major ($\theta_{\rm maj}$) and minor ($\theta_{\rm min}$) axes are calculated as the deconvolved full-width-half-maximum (FWHM) of the Gaussian fit for J0930+4614 and J1120+6711, while for J1643+3156 complex extended emission we manually measured a size of $\sim 1.9'' \times 2.3''$ within its $5\sigma$ contours. In Table \ref{tab:src_res} we report flux densities in column 1, and $\theta_{\rm maj}$ and $\theta_{\rm min}$ in columns 4 and 5, respectively.

Throughout the text, rest-frame luminosities at a given rest-frequency $\nu_\mathrm{rest}$ are calculated from the observed flux density $S_\mathrm{\nu,obs}$ at the observed frequency $\nu_\mathrm{obs}$ as:
\begin{equation}\label{eq:fluxtolum}
L_{\nu_\mathrm{rest}} = 4 \pi~D_L^2
\times \frac{S_{\nu_\mathrm{obs}}}{\nu_\mathrm{obs}^{\alpha}}
\times \left(\frac{\nu_\mathrm{rest}}{1+z}\right)^{1+\alpha}~~\mathrm{erg~s^{-1}}.
\end{equation}
We report in column 2 of Table \ref{tab:src_res} the targets' rest-frame luminosities at 144 MHz. For AGN A of J0930+4614 we use $\alpha = 0.08 \pm 0.1$ as derived from the 144 MHz -- 1.4 GHz flux densities (see Sec.~\ref{sec:J0930}), while for detected components of J1120+6711 and J1643+3156 we adopt a typical $\alpha = -0.7$ value, as well as for the upper limits to the undetected ones in all the targets.

The rest-frame $T_\mathrm{B}$ at $\nu_\mathrm{rest}$ can be calculated from  $S_\mathrm{\nu,obs}$ at $\nu_\mathrm{obs}$ as \citep{condon_1982, ulvestad_2005, njeri_2026}:
\begin{equation}\label{eq:tb}
T_\mathrm{B} = 1.22 \times 10^{3}(1+z) \left(\frac{S_\mathrm{\nu,obs}}{\text{mJy}}\right) \left(\frac{\mathrm{\nu_{obs}}}{\text{GHz}}\right)^{-2} \left(\frac{\theta_{\rm maj}\,\theta_{\rm min}}{\text{arcsec}^2}\right)^{-1} \text{K},
\end{equation}
Brightness temperature of the detected sources are reported in column 6 of Table \ref{tab:src_res}. The maximum rest-frame brightness temperature expected from SF processes can be calculated as \citep{condon_1992,morabito_2022}:
\begin{equation}\label{eq:tb_sf}
\begin{aligned}
&T_\mathrm{B,SF} = \left(\frac{T_\mathrm{e}}{\text{K}}\right) \left(1 - e^{-\tau}\right)
\left[1 + 10\,\left(\frac{\mathrm{\nu_{rest}}}{\text{GHz}}\right)^{0.1+\alpha}\right]\,~~ \mathrm{K} \\
&\text{with}\,~~\tau = \left(\frac{\nu_{\mathrm{rest}}}{\nu_0}\right)^{-2.1},
\end{aligned}
\end{equation}
where $T_{\rm e}$ is the electron temperature of the gas, and $\nu_0$ is the frequency at which the optical depth $\tau = 1$. Following \cite{morabito_2022}, in this work we adopt $T_{\rm e} = 10^4$ K and $\nu_0 = 3$ GHz. 

\begin{table*}
\caption{\label{tab:src_res} Results of fitting and measurements in the 144 MHz images. 
}
\centering
\begin{tabular}{cccccc}
\hline\hline
ID & $S_{\mathrm{144~MHz}}$ (mJy)& $L_{\mathrm{144~MHz}}$ ($10^{40}~\mathrm{erg/s}$)&  $\theta_{\rm maj}$ (\arcsec)& $\theta_{\rm min}$ (\arcsec) &  $T_\mathrm{B}$ ($10^6$ K)\\
(1) & (2) & (3) & (4) & (5) & (6)\\
\hline

\object{$\mathrm{J0930+4614~A}$} & $33\pm7$ & $50\pm10$ & $0.645 \pm 0.003$ & $0.455 \pm 0.001$ & $27 \pm 5$ \\
\object{$\mathrm{J0930+4614~B}$} & $<0.31$ & $<0.5$ &  &  & \\
\object{$\mathrm{J1120+6711~A}$} & $1.9\pm0.2$ & $2.9\pm0.3$ & $0.61 \pm 0.04$ & $0.51 \pm 0.03$ & $0.9 \pm 0.1$ \\
\object{$\mathrm{J1120+6711~B}$} & $0.8\pm0.2$ & $1.2\pm0.3$ & $0.6 \pm 0.2$ & $0.4 \pm 0.2$ & $0.4 \pm 0.2$ \\
\object{$\mathrm{J1643+3156~A}$} & $487\pm43$ & $86\pm 7$ & $\sim2.3$ & $\sim1.9$ & $10 \pm 1$ \\
\object{$\mathrm{J1643+3156~B}$} & $<0.5$ & $<0.08$ & & &  \\

\hline
\end{tabular}
\tablefoot{
(1) Source ID and component.
(2) 144 MHz flux density.
(3) 144 MHz luminosity, calculates assuming the measured $\alpha=0.08$ for J0930+4614 A and assuming $\alpha= -0.7$ for the rest of the components and upper limits.
(4) and (5) Major and minor axes. For J0930+4614 and 1120+6711 these are the components of the Gaussian fit, while for the extended complex emission of J1643+3156 we manually measured the size of the 5$\sigma$ emission region.
(6) Brightness temperature.
Upper limits are reported at the 5$\sigma$ level.
}
\end{table*}

Throughout the text, the spectral index $\alpha$ and its error between two flux densities $S_1$ and $S_2$ at frequencies $\nu_1 < \nu_2$ is computed as:
\begin{equation}\label{eq:spix_err}
\sigma_{\alpha} = \frac{1}{\left|\ln\left(\frac{\nu_2}{\nu_1}\right)\right|} \sqrt{\left(\frac{\sigma_{s_1}}{S_1}\right)^2 + \left(\frac{\sigma_{s_2}}{S_2}\right)^2}.
\end{equation}

\section{Analysis and discussion}


\label{sec:ind_sources}
\subsection{J0930+4614}
\label{sec:J0930}
In this source, we detected only AGN A ($S_{144~\mathrm{MHz}}=33 \pm7 $\,mJy). As mentioned in Sec. \ref{sec:intro}, ILT data are especially suited to identify AGN emission on the basis of the comparison between the source $T_\mathrm{B}$ and the maximum $T_\mathrm{B,SF}$ expected from SF. For example for AGN A we measured $T_\mathrm{B} = 2.7 \pm 0.5 \times 10^7$ K (Sec. \ref{sec:results}), that is $\approx 240 \times T_\mathrm{B,SF}$; thus, any source like J0930+4614 would be unambiguously classified as an AGN based only on its 144 MHz flux density.

AGN A is also detected at 1.4\,GHz in the NVAS image, for which we report a flux density of $S_{1.4~\mathrm{GHz}}=40 \pm4 $\,mJy. We note that the resolution of the VLA image is $\sim3.5''$, Thus, in principle, the emission could arise from the combined contribution of AGN A and B, given their separation of $\sim$1.5$''$. However, AGN B is not detected at 144 MHz nor in any image at 4.9, 8.5, 22.4\,GHz (whose resolution is $<0.9''$), thus we can fairly ascribe the entirety of the 1.4\,GHz emission to AGN A. The corresponding AGN spectral index in the 0.14--1.4 GHz range is $\alpha = 0.08 \pm 0.1$, fully consistent with a flat-spectrum core emission \citep{sotnikova_2021}. 
We can determine whether AGN A is classified as radio-loud (RL) or radio-quiet (RQ) by measuring the optical radio-loudness $R_\mathrm{O}$= $S_{\nu,5~\mathrm{GHz}}$/$S_{\nu,4400~\mathrm{\mathring{A}}}$ \citep{kellermann_1989}, where an AGN is RL if $R_\mathrm{O}>10$ \citep{jiang_2007}. Since both sources are resolved and detected in the optical spectra \citep{more_2016}, we can infer the AGN~A rest-frame luminosity at 4400 \AA\ ($L_{4400~\mathrm{\mathring{A}}}$) from the observed 5000 \AA\ flux density by assuming an average optical spectral index of $-0.5$ \citep{wang_2007}. As for rest-frame luminosity at 5 GHz $L_{5~\mathrm{GHz}}$, we derive it from $S_{1.4~\mathrm{GHz}}$ through Eq. \ref{eq:fluxtolum} by assuming the 0.14--1.4 GHz spectral index. We found $L_{4400~\mathrm{\mathring{A}}} = 1.7 \pm0.1 \times 10^{46}$ erg/s and $L_{5~\mathrm{GHz}} = 3.7 \pm0.4 \times 10^{43}$ erg/s, and $R_\mathrm{O} = 296 \pm 34$. This makes AGN A a clear RL AGN. The 1.4 GHz map sensitivity ($\sim1.4$ mJy at $5\sigma$) does not allow a meaningful discrimination between RQ/RL classifications for the undetected AGN B.




\subsection{J1120+6711}
\label{sec:J1120}
To our knowledge, this source is the first confirmed $z>1$ DAGN where both AGN are detected at LOFAR frequencies. In addition, it is also among the few objects of this class detected at any radio frequency, as well as one of the faintest \citep{mangat_2021, damato_2026}.
For AGN A we measured $T_\mathrm{B} = 9 \pm 1 \times 10^5$ K ($\sim 4 \times T_\mathrm{B,SF}$), while for AGN B we measured $T_\mathrm{B} = 4 \pm 2 \times 10^5$ K ($\sim 2 \times T_\mathrm{B,SF}$). This result highlights the potential of the ILT data to identify strong DAGN candidates purely from their radio flux density, even in the faintest cases. At GHz frequencies, it is currently impossible to disentangle SF from AGN contribution in $z>1$ DAGN candidates in the sub-mJy flux density regime with this technique \citep{mangat_2021}.
The rest-frame luminosity at 4400 \AA\ inferred from the optical spectra \citep{pindor_2006} is $L_{4400~\mathrm{\mathring{A}}} = 7.6 \pm0.4 \times 10^{45}$ erg/s for AGN A and $L_{4400~\mathrm{\mathring{A}}} = 7.5 \pm0.3 \times 10^{44}$ erg/s for AGN B. By assuming a range of radio spectral indexes $-0.7 < \alpha <0$, we found that $R_\mathrm{O} \sim 1$--9 for AGN A and $R_\mathrm{O} \sim 6$--40 for AGN B. That is, AGN A is RQ regardless of the spectral index, while AGN B can be RQ if $\alpha$ is sufficiently steep. Interestingly, we found that, for the same $\alpha$, the radio-brightest AGN is also the radio-quietest; besides the spectral index uncertainty, we note that the measured flux ratio between AGN A and B is $\sim2.4$ at 144 MHz, while it is $\sim10$ in the optical band. 

The source is also detected in X-rays in the DR14 of the XMM-Newton Serendipitous Source Catalog (4XMM; \citealt{webb_2020,traulsen_2020}), with a total flux of $F_{0.2 - 10 ~\mathrm{keV}} = 2.4 \pm 0.4 \times 10^{-13} ~\mathrm{erg/cm^2/s}$. By assuming a typical photon index $\Gamma=1.7$ \citep{bianchi_2009}, we modeled the X-ray spectrum with a power law using XSPEC \citep{arnaud_1996}, and derived a rest-frame $L_{2 - 10 ~\mathrm{keV}} \sim 1 \times 10^{45} ~\mathrm{erg/s}$. We found a column density $N_\mathrm{H}\sim5 \times 10^{20} ~\mathrm{cm^{-2}}$, that is, the source is unobscured (in agreement with the broad lines visible in the optical spectra of both AGN; \citealt{pindor_2006}). To infer the X-ray ratio of the two AGN, we converted the 4400 \AA\ luminosity to 2--10 keV luminosity using the bolometric correction $K_\mathrm{X} = L_\mathrm{BOL}/L_\mathrm{X}$ for Type 1 AGN found by \cite{duras_2020}:
\begin{equation}\label{eq:bol_corr}
\begin{aligned}
&K_\mathrm{X} = a~\left[ 1+ \left( \frac{\log(L_\mathrm{BOL}/L_\mathrm{\odot})}{b} \right)^c \right] \\
&\text{with}\,~~L_\mathrm{BOL} = 5.18~ L_\mathrm{4400~\mathring{A}},
\end{aligned}
\end{equation}
where $a= 12.76$, $b = 12.15$ and $c= 18.78$. We found $L_{\mathrm{2 - 10 ~keV}} \sim 7 \times 10^{44}$ erg/s for AGN A and $L_{\mathrm{2 - 10 ~keV}} \sim 2 \times 10^{43}$ erg/s for AGN B, respectively. The total $L_{\mathrm{2 - 10 ~keV}} \sim 7.2 \times 10^{44}$ erg/s is in good agreement with the $\sim10^{45}$ erg/s value directly measured from the XMM spectrum, considering the combined $\sim0.5$ dex scatter of the bolometric corrections and the optical slope uncertainty. This exercise suggests an even more extreme flux ratio between the two AGN in the X-rays ($\sim 35$) with respect to the optical band ($\sim$ 10) and, more importantly, the 144 MHz emission ($\sim$2.4). We note that in general different X-ray-to-radio emission ratios can correspond to different accretion efficiency regimes \citep{merloni_2003,panessa_2015,damato_2022}; while we do not attempt any physical interpretation based on the limited data of a single source, we emphasize that future studies based on multi-band resolved emission of a statistically meaningful sample of DAGN at different separations and redshifts, will be a valuable tool to investigate the accretion properties of this population. In addition, the J1120+6711 case demonstrates that with ILT data the selection and confirmation of DAGN can be highly efficient with respect to other bands that trace the inner accretion process (e.g., X-rays), where resolution limits and large flux ratios make it difficult to disentangle the two AGN contribution at less than few arcsec separation.






\subsection{J1643+3156}
\label{sec:J1643}

\begin{figure}[h!]
        \centering
\resizebox{\hsize}{!}
{\includegraphics[]{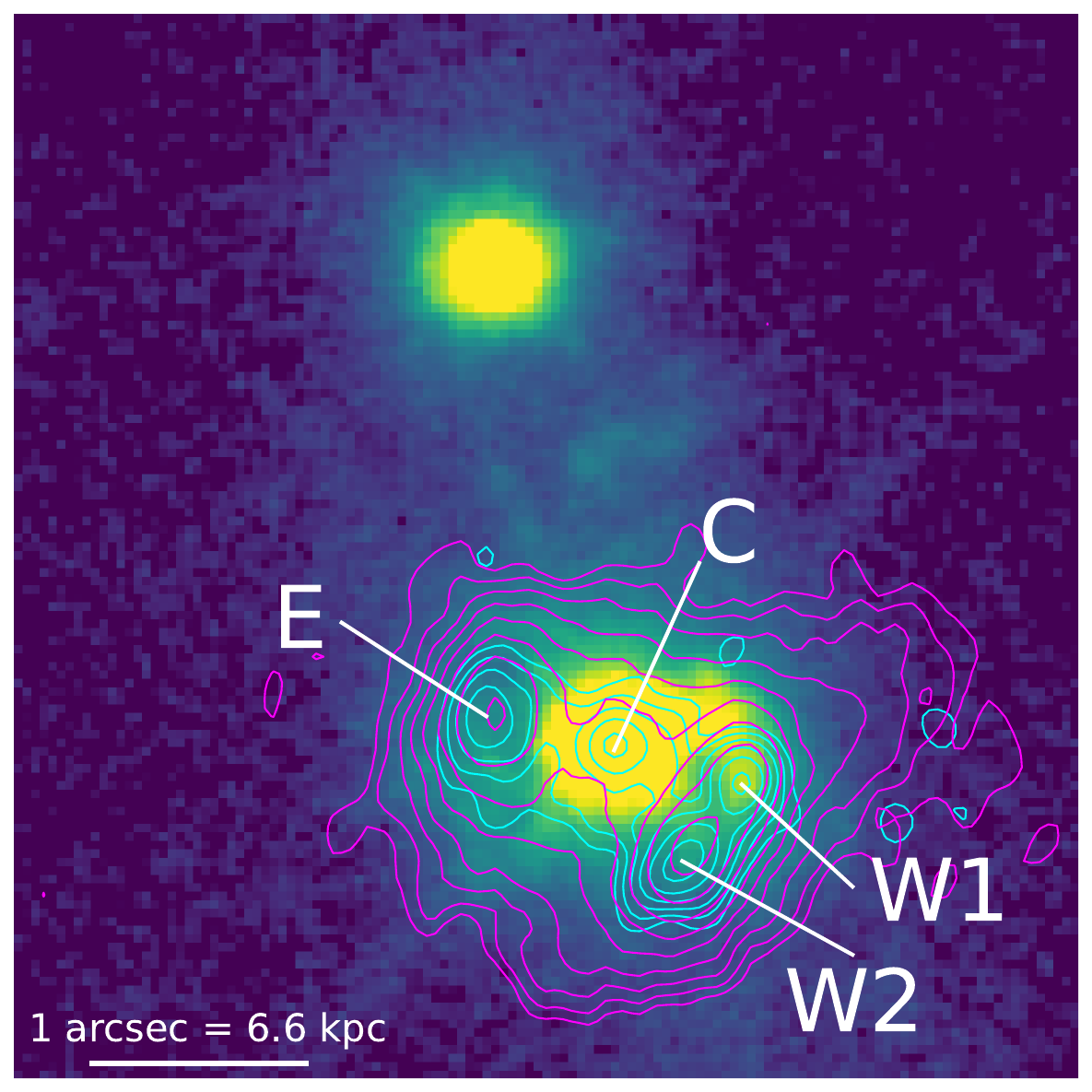}}

        \caption{HST/F814W image of J1643+3156. The 144 MHz ILT and 1.66 GHz MERLIN emission contours are overlaid in magenta and cyan, respectively, starting at the ${\sim}3 \sigma$ level and increasing with a $\sqrt{3}$ geometric progression. The four major components of radio emission are labeled in white (E, C, W1 and W2). The solid white line in the bottom-left corner indicates the angular and physical scale.}
        
        \label{fig:J1643_hst}
\end{figure}

AGN A component was initially classified as a CSS by \cite{brotherton_1999}, an evolutionary phase intermediate between the younger gigahertz peaked spectrum phase and large-scale radio sources \citep{fanti_1995,readhead_1996}. 
Since such young sources are commonly triggered by galaxy mergers (${\sim}$50\% show interaction signatures; \citealt{odea_1998,odea_2021}), their morphology can be used to investigate the merger-triggering scenario.

\begin{figure*}[h!]
        \centering
\resizebox{\hsize}{!}
{\includegraphics[]{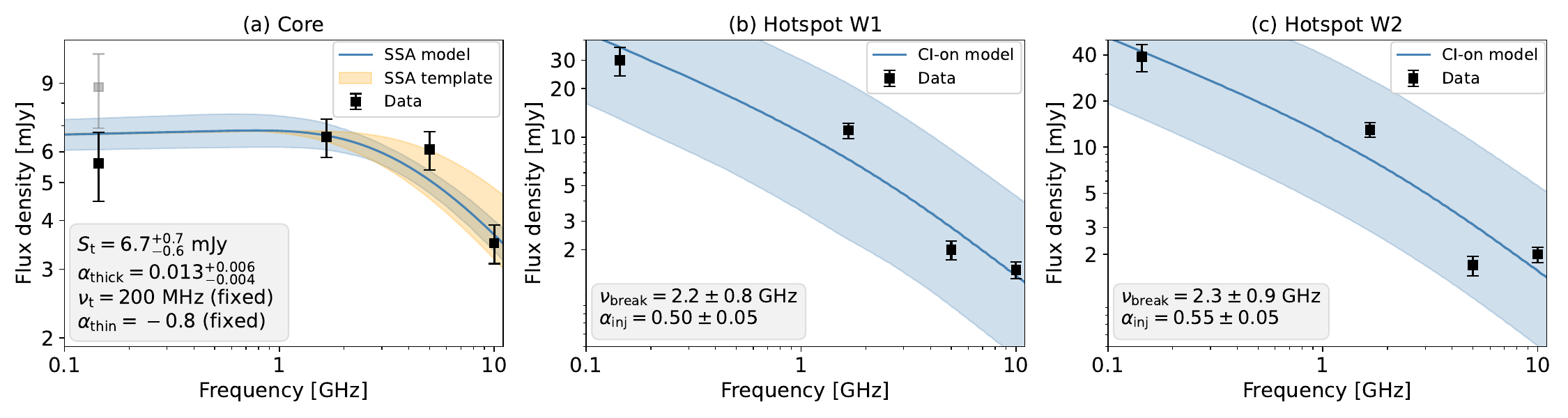}}

        \caption{Spectra of J1643+3156 components, namely the Core (a), the W1 hotspot (b) and the W2 hotspot (c). The black squares mark the measured peak flux densities at each frequency. The solid blue lines indicate the best model, while the light blue shades indicates uncertainty. The assumed (fixed) and best-fit parameters are reported in the light-grey boxes. Error bars and uncertainties are at the 1$\sigma$ level. For the Core, the light-gray square corresponds to the 144 MHz flux density measured in the lower resolution image, while the orange shade indicates the SSA template with $\nu_\mathrm{t}$ in the 100--300 MHz range.}
        
        \label{fig:J1643_fit}
\end{figure*}

J1643+3156 was observed by MERLIN at 1.66 GHz (cyan contours, Fig. \ref{fig:J1643_hst}). \cite{kunert-bajraszewska_2011} estimated a source age of few $\times 10^5$ yr from the E--W1 largest linear size assuming a lobe velocity of 0.1c, consistent with typical CSS values. Since this is an order of magnitude below the merger-induced tidal perturbation timescale (few $\times 10^7$ yr, derived from the projected distance and velocity offset of the companion), they concluded that jet formation was not directly triggered by the interaction, though the companion could still modulate the AGN duty cycle via prolonged (few $\times10^6$ yr) outbursts or jet precession. They proposed three scenarios: (1) the E--C--W1 structure traces past activity, while W2 is associated with the current jet; (2) jet precession, with W1 marking the initial direction and W2 the current one; or (3) jet interaction with the host-galaxy, which it is itself disturbed by the merger, as visible by its irregular optical morphology and the emission clumps in the region between the two AGN hosts (Fig. \ref{fig:J1643_hst}, see \citealt{kunert-bajraszewska_2011} and \citealt{martel_2005} for optical image discussion). In this case, the entire radio structure is confined within the galaxy, with the W1--W2 hotspots arising from the jet forcing its way through the inter-stellar medium (ISM). We can now test these scenarios using the radio spectrum from 144 MHz to 10 GHz (Fig. \ref{fig:J1643_fit}, Table \ref{tab:J1643_peaks}), measured at each component's emission peak to minimize contamination from diffuse structure.

The core spectrum (Fig. \ref{fig:J1643_fit}a) is flat between 1.66 and 5 GHz ($\alpha =-0.07 \pm 0.14$), as typically observed for compact nuclear emission in the self-synchrotron absorption (SSA) regime \citep[e.g.,][]{jones_1974}. Above 5 GHz, it steepens sharply ($\alpha =-0.8 \pm 0.2$), marking the transition to the optically thin regime \citep[e.g.,][]{mikhailov_2026}. The 144 MHz flux density initially appeared to lie above this plateau (light-gray point); a higher resolution map (Fig. \ref{fig:J1643_highres}) confirmed this was due to residual jet contamination, and we adopted the high-resolution value in the fit. We fitted an SSA model (characterized by the flux density normalization $S_\mathrm{t}$, the spectral indices in the optically thin and thick regimes $\alpha_\mathrm{thin}$ and $\alpha_\mathrm{thick}$, and the turnover frequency $\nu_\mathrm{t}$; see Eq. \ref{eq:SSA}). We fixed $\alpha_\mathrm{thin} = -0.8$ (i.e, the $\alpha$ value between 5 and 10 GHz) and $\nu_\mathrm{t}=200$ MHz, that is the median value for CSS sources of comparable size and redshift \citep{odea_1998}. To test the impact of the latter assumption, we varied $\nu_\mathrm{t}$ over the 100--300 MHz scatter typical for CSS (orange shade, Fig. \ref{fig:J1643_fit}a), while keeping the other parameters fixed to the previous fit assumptions and best-fit values. In any case, the result is a well-constrained, flat-spectrum ($\alpha_\mathrm{thick}\sim0$) SSA model, in agreement with the data within $\sim1\sigma$. Our analysis and tests indicate that the core spectrum is consistent with that expected for a typical CSS nucleus of similar size and redshift.

W1 and W2 have comparable flux densities at all frequencies, suggesting a common evolutionary history (Fig. \ref{fig:J1643_fit}b,c). We fitted both with \texttt{syncrofit} \citep{quici_2022}, assuming a continuous-injection model characterized by the break frequency $\nu_\mathrm{break}$ and injection index $\alpha_\mathrm{inj}$ (CI-on; \citealt{kardashev_1962,komissarov_1994}; see Appendix \ref{ap:J1643_analysis}). The best-fit models confirm a shared evolutionary history between the two hotspots. Following \cite{turner_2018} and assuming a magnetic field $B =100$--200 $\mu$G, typical of CSS hotspots \citep{odea_1998,murgia_2003}, we estimate an age of a few $\times 10^5$ yr for W1 and W2, consistent with \cite{kunert-bajraszewska_2011}; even for a low $B=50~\mu$G, the age is at most few $\times 10^6$ yr, still an order of magnitude below the merger timescale. Notably, W1 coincides with an irregular optical arm of the host galaxy (Fig. \ref{fig:J1643_hst}), suggesting it sits in a high-density ISM region. The higher-resolution image reveals a similarly complex and bent morphology on the E side (Fig. \ref{fig:J1643_highres}).

In summary our analysis of AGN A in J1643+3156 has shown that: \textit{i)} the nucleus shows SSA emission typical of a CSS of similar size and redshift; \textit{ii)} W1 and W2 share comparable spectra, particle-acceleration histories, and radiative ages, at least an order of magnitude below the merger perturbation timescale; \textit{iii)} the extended emission is confined within the host galaxy -- itself disturbed by the ongoing merger -- with one hotspot coincident with a likely high-density ISM region. We therefore favor interaction with the host ISM over merger-regulated AGN activity as the origin of the source's complex morphology. The new ILT data were essential in constraining the low-frequency spectrum and enabling this analysis.

For AGN B, we obtained a $5\sigma$ upper limit of $\sim0.5$ mJy, with non-detections at 1.6, 5, and 10 GHz. \cite{brotherton_1999} reported a tentative $8.2$ GHz detection ($0.4$ mJy, $\sim4\sigma$). Combined with our non-detections, this would imply a flat or inverted spectrum up to high frequency, that can not be excluded for compact nuclear emission. However, inspection of the 8.2 GHz image publicly available in the NVAS (project ID: AB863) shows that the AGN B position coincides with a side lobe of AGN A, suggesting that the reported  detection may instead be an imaging artifact.

\section{Future perspectives}
\label{sec:fut_per}

\subsection{DAGN detection rate}
\label{sec:det_rate}
The detection of 6 sources in the 6\arcsec\ resolution LoTSS DR2 catalog, out of 25 sources in its footprint, corresponds to a detection rate of $24$\%. A detection rate of $\sim$24\%  is also found in the most recent LoTSS data realease (DR3; \citealt{shimwell_2026}): among the 92 DAGN of our initial sample, 63 fall in the DR3 footprint, and 15 are detected at $>5\sigma$ and associated with the DR3 source catalogue ($>$13.6M sources distributed over 88\% of the northern sky). 
We note that all our objects come from optically selected samples of dual or lensed AGN candidates that have been subsequently confirmed as DAGN through spectroscopic follow-ups; thus, in order to investigate whether DAGN exhibit an enhanced radio detectability in the LoTSS with respect to single AGN, we can compare this detection rate with that of optically selected and spectroscopically confirmed AGN in the combined SDSS DR16 \citep{lyke_2020} and Dark Energy Spectroscopic Instrument (DESI) DR1 \citep{desi_collab_2026}. Among the $\sim$1.15M spectroscopically confirmed AGN at $z>0.3$ in the SDSS+DESI catalog, $\sim$520k and $\sim$1M objects are in the DR2 and DR3 LoTSS footprints, respectively. In agreement with the DAGN sample, we found that also single AGN present a comparable detection rate between the two releases, that is $\sim$10\% for DR2 and $\sim$9\% for DR3 (i.e., a factor of ${\sim}2.5$ lower than DAGN). In order to assess the significance of the detection excess of DAGN, we can perform a one-sided binomial test; for DR2, the test returns a p-value of $P(X \geq 6, ~n=25~|~ p= 0.10)\sim0.03$, that is, we exclude at $\sim97$\% confidence the possibility of obtaining 6 or more detections purely by chance. Notably, for DR3 the detection excess with respect to single AGN is much more significant: $P(X \geq 15, ~n=63~|~ p= 0.09)\sim 4\times 10^{-4}$. 
If confirmed, the remarkable detection rate of DAGN could be explained by either a difference in the intrinsic physical properties of each AGN in the pair compared to isolated AGN, possibly triggered by the merging events (as predicted by a number of hydrodynamical simulations; see \citealt{van_wassenhove_2012,capelo_2015}), or the blending effect in 6\arcsec resolution LOFAR images, where the combined emission from two AGN contributes to the total observed flux density. While the detection of both AGN already in one out of three cases of the pilot sample analyzed here (i.e., J1120+6711; Sec. \ref{sec:J1120}) points towards the latter explanation, recent X-ray analyses found that AGN activity can be already enhanced when the DAGN have a distance of several tens of kpc \citep{treister_2012,koss_2012,derosa_2023, battistini_2026}. Since, according to the AGN fundamental plane, the X-ray and radio emission are tightly related for a large range of BH masses \citep{merloni_2003,panessa_2015,damato_2022}, one may also expect an enhanced radio activity in DAGN. In either case, our result suggests that radio observations provide an efficient method to detect DAGN emission and to investigate their properties. 
 
Testing the aforementioned hypotheses requires a larger sample with high-resolution ILT imaging follow-up in the LoTSS. This is beyond the scope of this work and will be addressed in a future work, but as a preliminary test we produced 6\arcsec~LOFAR cutouts from the public LoTSS images around all the 25 sources in DR2, and found that 21 (unresolved) DAGN are detected at $\geq3 \sigma$. This means that in $\sim84$\% of DAGN systems at least one component is detected. These DAGN span a wide and uniformly distributed range of projected separations ($2.5 ~\mathrm{kpc} < r_\mathrm{p} < 30~\mathrm{kpc}$) and redshifts ($0.6 < z < 3.1$). Their re-imaging at 0.3$''$ resolution will provide the first statistically meaningful characterization of DAGN emission at 144 MHz, and in general one of the largest samples to date of high-redshift ($z > 0.5$) DAGN studied at radio frequencies (see \citealt{damato_2026} and references therein). The sample can be further expanded by exploiting the most recent DR3 of the LoTSS. If in $\sim$1/3 of such systems both AGN will be detected in ILT imaging, such as we found in the small pilot sample analyzed in this work (Sec. \ref{sec:J1120}), the luminosity and separation distributions, the radio-detected fraction of optical emitters, radio-loudness, and many other properties can be investigated in a wide range of redshift and at the smallest known separation scales \citep{mannucci_2023}. The above quantities and their distributions are predicted by several galaxy evolution models, and these data will allow us to put observational constraints to different evolutionary scenarios (e.g. \citealt{tremmel_2017,chen_2023_sim,shen_2023,volonteri_2003,volonteri_2022,dong_paez_2023}). 

\subsection{DAGN selection and comparison with SKA}

\begin{figure*}[h!]
        \centering
\resizebox{\hsize}{!}
{\includegraphics[width=0.5\textwidth]{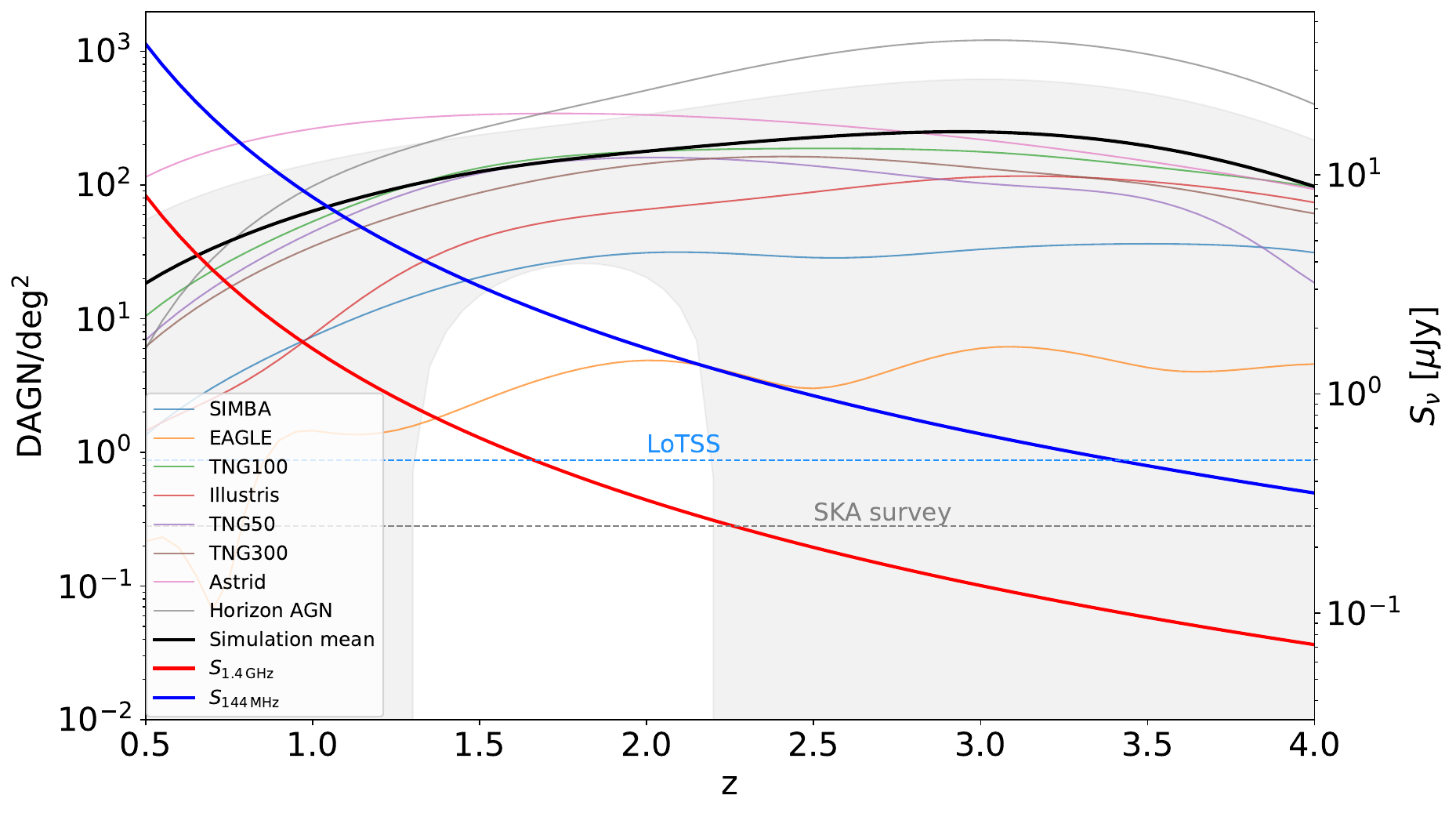}}

        \caption{Number of DAGN per $\mathrm{deg}^2$ (left $y$-axis) as a function of redshift, for a number of hydrodynamical simulations. The thick black solid line is the mean of the simulations, while the gray area is the standard deviation. The thick blue (red) solid line is the expected flux density (right $y$-axis) at 144 MHz (1.4 GHz) for an $L_\mathrm{bol} > 10^{43}$ erg s$^{-1}$ AGN. The gray dashed horizontal line is the $5\sigma$ detection threshold for a proposed SKA reference survey of $\sim 1 ~\mathrm{deg}^2$ at 1.4 GHz, while the light-blue dashed horizontal line is the $5\sigma$ detection threshold of the LoTSS. }
        
        \label{fig:DAGN_prediction}
\end{figure*}

Building on this pilot sample, ILT images offer a powerful tool for identifying new DAGN candidates directly in the radio band -- currently one of the major challenges in DAGN selection -- and potentially unveiling the high-$z$, heavily obscured population missed by standard optical/X-ray selection \citep{damato_2026}. We showed that ILT imaging can detect a DAGN down to a few mJy total emission out to $\sim$0.5 deg from the delay calibrator, meaning any LOFAR-detected source in this region can be quickly re-imaged at 0.3$''$ resolution via a simple phase-shift of the calibration solutions followed by cleaning. Applying this to the 21 DAGN in the DR2 sample detected at $>3\sigma$ yields $\sim5.3~\mathrm{deg}^2$ area; combined with the existing ILT LoTSS deep fields (Lockman Hole, ELAIS-N1, and Boötes; \citealt{morabito_2022,de_jong_2024,escott_2026}), this extends the candidate search to $\sim25.5~\mathrm{deg}^2$. As shown in Sec. \ref{sec:ind_sources}, ILT 144 MHz emission is well suited to identify AGN emission from flux density alone, disentangling it from SF contamination even in the faintest sources via $T_\mathrm{B}$ selection. The combined LoTSS high-resolution area will thus enable immediate identification of true AGN emission, with close-separation point sources emerging as DAGN candidates. This can be refined down to sub-arcsec separation with optical methods such as the Gaia multi-peak (GMP) technique \citep{mannucci_2022,mannucci_2023}, which searches for multiple peaks in quasar light profiles in the \textit{Gaia} archive; identifying optical continuum cospatial with the radio point sources will allows exclusion of multiple jet hotspots, strengthening the radio-selected candidates. Commonly to all selection methods, final confirmation will require spectroscopic follow-up, achievable at sub-arcsec separation via adaptive-optics-assisted integral field unit observations (e.g., \citealt{mannucci_2023,ciurlo_2023,chen_2023_vodka,scialpi_2024,scialpi_2025}).

Constraining the DAGN rate is an essential empirical input for super-massive black hole binary population models, improving predictions for the merger events LISA will observe \citep{Amaro-Seoane_2023,singh_2026}. Deep radio surveys are especially effective at uncovering low- to moderate-luminosity AGN missed by photometric classification \citep{smolcic_2017,damato_2022}; however, estimates of DAGN radio properties still rely heavily on empirical scaling relations with optical/X-ray emission derived from single-AGN populations, and on simplified assumptions \citep{dong_paez_2023}. While SKA will be essential for strong constraints at GHz frequencies, LoTSS already offers comparable or better sensitivity, resolution, and sky coverage at lower frequencies, providing an early opportunity to test these predictions.

Following \cite{damato_2026}, Fig. \ref{fig:DAGN_prediction} shows the number of DAGN per $\mathrm{deg}^2$ from the hydrodynamical simulation predictions of \cite{puerto_sanchez_2025}, including AGN as faint as $L_\mathrm{bol} > 10^{43}$ erg s$^{-1}$ and host masses down to $M_\ast = 10^9~\mathrm{M_\odot}$. The predicted DAGN density spans $\sim$3 orders of magnitude (few to several thousand per $\mathrm{deg}^2$) at any redshift. Since no direct theoretical link exists between bolometric and radio luminosity, \cite{damato_2026} converted $L_\mathrm{bol} = 10^{43}$ erg s$^{-1}$ to X-ray luminosity following \cite{puerto_sanchez_2025}, then to 1.4 GHz radio luminosity using the empirical correlation of \cite{damato_2022} (valid for $0 \lesssim z \lesssim 3$), and finally to observed flux density $S_{1.4~\mathrm{GHz}}$ assuming $\alpha=-0.7$ (solid red line). Assuming the same spectral index, we derived the observed $S_{144~\mathrm{MHz}}$ (solid blue line). This highlights LOFAR's effectiveness at detecting DAGN: at the SKA reference-survey sensitivity limit \citep{prandoni_2015} (gray dashed line), such an AGN would be detected at 1.4 GHz up to $z\sim2.2$, while the current LoTSS sensitivity (light-blue dashed line) reaches $z\sim3.5$. This work lays the groundwork for DAGN selection in the upcoming iLoTSS survey, which will provide homogeneous ILT 0.3$''$-resolution imaging of the northern sky and a benchmark for future SKA surveys.

\section{Conclusions}
\label{sec:concl}

We presented the first results of a pilot project aimed at studying DAGN at low radio frequencies (144 MHz). By cross-matching a sample of 92 spectroscopically confirmed systems at $z>0.3$ and projected separation $r_\mathrm{p}< 30$ kpc with the LoTSS DR2 catalog, we found that six sources are detected at $>5\sigma$ significance in the 6\arcsec\ resolution LOFAR images. As a proof of concept, we reprocessed LoTSS observations of three of these objects spanning $0.59<z<2.39$ and a broad range of observational conditions, including the ILT baselines to achieve an image resolution of $\sim 0.3''$. By analyzing their emission, we found the following main results:

\begin{itemize}

\item We successfully produced high-quality sub-arcsec resolution ILT images for all three targets. In one source (J1120+6711), we detected both AGN components at $>5\sigma$ significance, while in the other two we detect only one AGN. To our knowledge, J1120+6711 is the first confirmed $z>1$ DAGN for which both nuclei are detected at LOFAR frequencies, and it is one of the faintest ever detected at any radio frequencies. This result highlights the capability of low-frequency radio observations to efficiently identify and characterize high-$z$ kpc-scale separation DAGN.

\item The measured $T_\mathrm{b}$ of J1120+6711 and J0930+4614 components confirmed that ILT 144 MHz imaging can effectively distinguish AGN-powered radio emission from SF processes, enabling the identification of strong DAGN candidates directly from radio data. This approach is particularly promising for uncovering obscured systems that may be missed by optical and X-ray selection techniques. 

\item In the case of J1643+3156, where only one AGN is detected, the new ILT images allowed us to trace the extended synchrotron emission. By constraining the low-frequency radio spectrum of the source, we demonstrated that its radio properties are consistent with those of a young CSS source. Its complex morphology is most likely shaped by interactions between the radio jets and the ISM of the host galaxy, rather than by direct influence from the companion galaxy. This highlights the value of these observations in providing valuable physical insight and distinguishing among different evolutionary scenarios.

\item Considering the full sample of spectroscopically confirmed DAGN, we find a remarkable detection rate of $\sim$24\% in both 6\arcsec\ resolution LoTSS DR2 and DR3, compared to only $\sim$9--10\% of spectroscopically confirmed single AGN with matching selection criteria. A one-sided binomial test yields $P\sim 4\times 10^{-4}$ for the null hypothesis in the DR3, demonstrating a statistically significant excess of radio detections among DAGN. This excess may reflect enhanced AGN activity during galaxy interactions, the combined contribution of two blended active nuclei, or both effects. Regardless of the underlying cause, the result poses low-frequency radio surveys as a promising method for efficiently detecting DAGN.

\item The LoTSS data will allow the detection of DAGN emission up to $z \sim 3.5$ at 144 MHz, compared to $z \sim 2.2$ achievable at 1.4 GHz in SKA reference surveys, for an AGN with $L_\mathrm{bol} = 10^{43}~\mathrm{erg~s^{-1}}$. This highlights the unique role of LOFAR in building statistically significant DAGN samples and testing the predictions of hydrodynamical simulations.
\end{itemize}

\begin{acknowledgements}
This research made use of the LOFAR-IT computing infrastructure supported and operated by INAF, including the resources within the PLEIADI special ``LOFAR'' project by USC-C of INAF, and by the Physics Dept. of Turin University (under the agreement with Consorzio Interuniversitario per la Fisica Spaziale) at the C3S Supercomputing Centre, Italy. LOFAR is the Low Frequency Array designed and constructed by ASTRON. It has observing, data processing, and data storage facilities in several countries, which are owned by various parties (each with their own funding sources), and which are collectively operated by the LOFAR ERIC under a joint scientific policy. The LOFAR resources have benefited from the following recent major funding sources: CNRS-INSU, Observatoire de Paris and Université d'Orléans, France; BMFTR, MKW-NRW, MPG, Germany; Science Foundation Ireland (SFI), Department of Business, Enterprise and Innovation (DBEI), Ireland; NWO, The Netherlands; The Science and Technology Facilities Council, UK; Ministry of Science and Higher Education, Poland; The Istituto Nazionale di Astrofisica (INAF), Italy. This research made use of the Dutch national e-infrastructure with support of the SURF Cooperative (e-infra 180169) and the LOFAR e-infra group. The Jülich LOFAR Long Term Archive and the German LOFAR network are both coordinated and operated by the Jülich Supercomputing Centre (JSC), and computing resources on the supercomputer JUWELS at JSC were provided by the Gauss Centre for Supercomputing e.V. (grant CHTB00) through the John von Neumann Institute for Computing (NIC). This research made use of the University of Hertfordshire high-performance computing facility and the LOFAR-UK computing facility located at the University of Hertfordshire and supported by STFC [ST/P000096/1], and of the Italian LOFAR-IT computing infrastructure supported and operated by INAF, including the resources within the PLEIADI special "LOFAR" project by USC-C of INAF, and by the Physics Department of Turin university (under an agreement with Consorzio Interuniversitario per la Fisica Spaziale) at the C3S Supercomputing Centre, Italy. This research is part of the project LOFAR Data Valorization (LDV) [project numbers 2020.031, 2022.033, and 2024.047] of the research programme Computing Time on National Computer Facilities using SPIDER that is (co-)funded by the Dutch Research Council (NWO), hosted by SURF through the call for proposals of Computing Time on National Computer Facilities. 
We acknowledge the support of the computing centre of INAF-Osservatorio Astronomico di Trieste, under the coordination of the CHIPP project \citep{taffoni_2020,bertocco_2020}.
We acknowledge financial contribution from INAF Large Grant "The Quest for dual and binary massive black holes in the gravitational wave era", (Bando Ricerca Fondamentale INAF 2024).
We acknowledge financial support from the ASI agreement n. 2024-36-HH.1-2025 and 2025-29-HH.0.
Image credit: NRAO/VLA Archive Survey, (c) 2005-2007 AUI/NRAO.

EDR acknowledges support by the Deutsche Forschungsgemeinschaft (DFG).

AJ acknowledges financial support from the Polish National Science Center under grant 2023/50/A/ST9/00527.

CS acknowledges financial support by the Italian Ministry of University and Research (grant FIS2023$-$01611, CUP C53C25000300001) and by the INAF (through “Ricerca Fondamentale 2024”, RSN4 mini-grant 1.05.24.07.04).

GV acknowledges financial support from the Italian National Institute for Astrophysics (INAF) under the IAF - Astrophysics Fellowships in Italy grant CUP C59J21034720001 - ``AD MAJORA''.

EB acknowledges the support of the “Ricerca Fondamentale 2024” INAF program (GO grant ``A JWST/MIRI MIRACLE: Mid-IR Activity of Circumnuclear Line Emission'' and RSN1 mini-grant 1.05.24.07.01). 

AP acknowledges support from ANID/FONDECYT/Postdoctorado 3260848.
\end{acknowledgements}

%
\bibliographystyle{aa} 
\bibliography{bibliography} 

\begin{appendix}
\section{Additional details of J1643+3156 analysis}
\label{ap:J1643_analysis}

\begin{table}[ht!]
\caption{\label{tab:J1643_peaks} Peak flux densities at different frequencies for J1643+3156}
\centering
\resizebox{\hsize}{!}{
\begin{tabular}{ccccc}
\hline \hline
Frequency & $S_{\mathrm{peak, C}}$ & $S_{\mathrm{peak, W1}}$ & $S_{\mathrm{peak, W2}}$ & Ref. \\
(1) & (2) & (3) & (4) & (5) \\
\hline
0.14  & $5.6 \pm 1.7$ & $30.0 \pm 6.0$ & $39.0 \pm 7.8$ &  This work\\
1.6   & $6.5 \pm 0.7$ & $10.8 \pm 1.2$ & $12.8 \pm 1.4$ & 1\\
5.0   & $6.1 \pm 0.7$ & $1.9 \pm 0.3$  & $1.7 \pm 0.2$  & 1 \\
10.0  & $3.5 \pm 0.4$ & $1.5 \pm 0.2$  & $2.0 \pm 0.2$  & 2 \\
\hline
\end{tabular}
}
\tablefoot{(1) Observing frequency in GHz. (2 -- 4) Peak flux densities of component C, W1 and W2 in mJy, respectively. (5) Reference for the measurement: 
1 is \cite{kunert-bajraszewska_2011} and 2 is \cite{Maithil_2020}.}
\end{table}

\subsection{The core}

\begin{figure}[ht!]
        \centering
\resizebox{\hsize}{!}
{\includegraphics[]{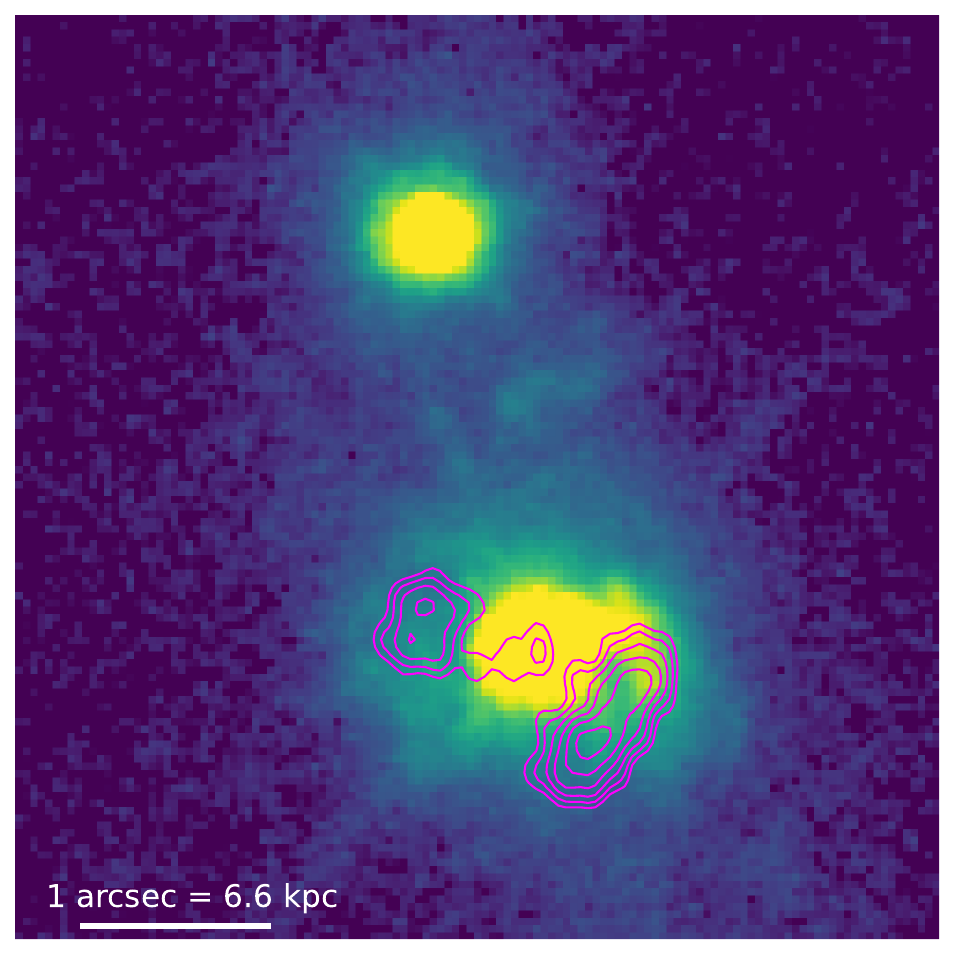}}

        \caption{HST/F814W image of J1643+3156, with superimposed 144 MHz LOFAR contours in magenta. The LOFAR contours, starting at 4.5 mJy and increasing with a $\sqrt{2}$ geometric progression, are drawn from an image obtained with weighting parameter Briggs $= -2$ (corresponding to uniform weighing) and a $uv$-cut $>60 \mathrm{k\lambda}$, in order to maximize the resolution and minimize contribution of extended emission. }
        
        \label{fig:J1643_highres}
\end{figure}

We used peak emission to trace the inner AGN regions while minimizing contamination from jet-base extended emission, clearly detected towards the West in the highest-resolution 10 GHz image \citep{Maithil_2020}; this is especially important for the ILT image, given its higher resolution and sensitivity to extended emission. The 144 MHz point (light-gray point in Fig. \ref{fig:J1643_fit}a) lies above the 1.66--5 GHz plateau ($\alpha=-0.07\pm0.14$), though the 0.14--1.66 GHz index ($\alpha=-0.12\pm0.09$) is still consistent with 0 within ${\sim}1\sigma$ and with the standard flat-spectrum range $-0.5 \leq \alpha \leq 0$ \citep[e.g.,][]{sotnikova_2021}. We ascribe this discrepancy to residual jet-extended emission in the 144 MHz peak pixel, and tested this by producing a higher-resolution LOFAR image (Briggs robust $=-2$, i.e. uniform weighting, with a $uv$-cut $>60~\mathrm{k\lambda}$; Fig. \ref{fig:J1643_highres}).

In this image the core peak is $5.6\pm1.2$ mJy, $\sim36\%$ lower than the lower-resolution value (versus at most $\sim12\%$ for the other components), and reveals a jet connecting the core to component E; we therefore adopt this high-resolution value for C in the analysis. Above 5 GHz the flux density drops steeply ($\alpha=-0.8\pm0.2$), consistent with the transition from the optically thick to the optically thin regime typically seen in compact sources \citep[e.g.,][]{mikhailov_2026}.

SSA emission from a power-law electron distribution in a magnetic field is described by \citep{turler_1999}:
\begin{equation}\label{eq:SSA}
S_\nu = S_\mathrm{t} \left( \frac{\nu}{\nu_\mathrm{t}} \right)^{\alpha_{\mathrm{thick}}}
\frac{1 - \exp\left(-\tau_\mathrm{t} \left( \frac{\nu}{\nu_\mathrm{t}} \right)^{\alpha_{\mathrm{thin}} - \alpha_{\mathrm{thick}}} \right)}
{1 - \exp(-\tau_\mathrm{t})}
\end{equation}
where $\tau_\mathrm{t}=3/2 (\sqrt{1-(8\alpha_\mathrm{thin}/3\alpha_\mathrm{thick})}-1)$ is the optical depth at the turnover. Given the few spectral points, $\alpha_\mathrm{thick}$ and $\nu_\mathrm{t}$ are degenerate if left both free during the fit; we therefore fixed $\nu_\mathrm{t}=200$ MHz (the median value for CSS sources of comparable size and redshift; \citealt{odea_1998}) and $\alpha_\mathrm{thin}=-0.8$ (from the 5--10 GHz data, consistent with the typical value ${\sim}-0.7$; \citealt{zajacek_2019}), and fitted the remaining parameters via Markov chain Monte Carlo (MCMC) with {\sc{emcee}} (\citealt{foreman_2013}), obtaining a SSA fit in ${\sim}1\sigma$ agreement with the data. To test the impact of the most critical assumption on $\nu_\mathrm{t}$, we repeated this with $\nu_\mathrm{t}$ spanning the 100--300 MHz scatter found for comparable CSS sources \citep{odea_1998}, while keeping $\alpha_\mathrm{thin}=-0.8$ and $S_\mathrm{t}$ fixed to the best fit values; the resulting spread (orange shade, Fig. \ref{fig:J1643_fit}a) still matches the data within 1$\sigma$.

\subsection{W1 and W2 hotspots}

The 0.14--1.66 GHz spectral index is $\alpha=-0.41\pm0.09$ (W1) and $-0.45\pm0.09$ (W2), mildly flat, steepening to $\alpha=-1.6\pm0.1$ and $-1.8\pm0.2$ in the 1.66--5 GHz range (a spectral break commonly observed in CSS lobes/hotspots; \citealt{odea_1998,murgia_1999,murgia_2003,odea_2021}). The break frequency $\nu_\mathrm{break}$ traces the electron energy distribution and radiative age of the plasma, while the 144 MHz point constrains the pre-ageing injection index $\alpha_\mathrm{inj}=-(p-1)/2$, related to the initial electron distribution $N(E)\propto E^{-p}$. Thus, the large frequency coverage of these objects allowed us to break the $\nu_\mathrm{break}$--$\alpha_\mathrm{inj}$ degeneracy despite the few points. We fitted W1 and W2 peak fluxes (Table \ref{tab:J1643_peaks}) with \texttt{syncrofit} \citep{quici_2022}, using peak emission to minimize diffuse contamination since standard ageing models assume a uniform field and a common injection history.

Classical impulsive-injection ageing models -- Kardashev--Pacholczyk \citep[KP;][]{kardashev_1962,pacholczyk_1970} and Jaffe--Perola \citep[JP;][]{jaffe_1973} -- describe an electron population injected at $t_\mathrm{i}$ and steepening above a break frequency as it ages radiatively. These models are not well suited for actively accelerating sources, and we instead adopt the CI framework \citep{kardashev_1962,komissarov_1994}, in which electrons are supplied continuously from $t_\mathrm{i}$ to $t_\mathrm{off}$. We specifically adopt the CI-on model, in which particle injection is still ongoing at the source age $t_\mathrm{age}$. This choice is motivated by the SSA core spectrum, which indicates an active CSS.  Model errors are estimated from 500 Monte Carlo iterations.

\onecolumn
\section{List of selected DAGN}
The pilot sample analyzed in this work is drawn from a parent sample of 92 DAGN at $z>0.3$ with projected physical separations $r_\mathrm{p}<30$ kpc that, to the best of our knowledge, had been spectroscopically confirmed as of January 2025. In Table \ref{tab:full_sample} we list all these objects, along with their coordinates, redshifts, and angular and physical separations.

\begin{longtable}[ht!]{cccccc c}

\caption{Summary of the targets. From left to right: Source ID, right ascension, declination, redshift, projected angular separation, projected physical separation, reference of the spectrum.}\\
\label{tab:full_sample} \\
\hline\hline
ID & RA(J2000) & DEC(J2000) & $z$ & $\theta_\mathrm{_{p}}$ ($''$) & $r_\mathrm{_{p}}$ (kpc) & Ref. \\
\hline
\endfirsthead
\caption{continued.}\\
\hline
ID & RA(J2000) & DEC(J2000) & $z$ & $\theta_\mathrm{_{p}}$ ($''$) & $r_\mathrm{_{p}}$ (kpc) & Ref. \\
\hline
\endhead
\hline
\endfoot
\object{$\mathrm{J0027+0232}$} & $00^h 27^m 02^s.86$ & $+02^{\circ} 32' 15''.00$ & 2.020 & 2.72 & 22.7 & \citet{lemon_2020} \\
\object{$\mathrm{J0052-3045}$} & $00^h 52^m 37^s.98$ & $-30^{\circ} 45' 52''.24$ & 1.455 & 0.43 & 3.6 & \citet{scialpi_2024} \\
\object{$\mathrm{J0058-6120}$} & $00^h 58^m 17^s.11$ & $-61^{\circ} 20' 04''.55$ & 1.322 & 3.03 & 25.4 & \citet{aguita_2018} \\
\object{$\mathrm{J0103-2753}$} & $01^h 05^m 34^s.73$ & $-27^{\circ} 36' 57''.92$ & 0.834 & 0.29 & 2.2 & \citet{junkkarinen_2001} \\
\object{$\mathrm{J0118-0104}$} & $01^h 18^m 12^s.02$ & $-01^{\circ} 04' 42''.60$ & 0.739 & 1.74 & 12.7 & \citet{lemon_2020} \\
\object{$\mathrm{J0118-3115}$} & $01^h 18^m 40^s.31$ & $-31^{\circ} 15' 42''.52$ & 1.740 & 1.00 & 8.5 & \citet{lemon_2020} \\
\object{$\mathrm{J0120-4354}$} & $01^h 20^m 06^s.38$ & $-43^{\circ} 54' 40''.8$ & 1.910 & 0.84 & 7.1 & \citet{aguita_2018} \\
\object{$\mathrm{J0122+0358}$} & $01^h 22^m 23^s.77$ & $+03^{\circ} 58' 37''.25$ & 1.690 & 1.71 & 14.5 & \citet{lemon_2020} \\
\object{$\mathrm{J0127-1441}$} & $01^h 27^m 08^s.49$ & $-14^{\circ} 41' 19''.00$ & 1.760 & 2.96 & 25.0 & \citet{lemon_2018} \\
\object{$\mathrm{J0139+3526}$} & $01^h 39^m 33^s.33$ & $+35^{\circ} 26' 11''.7$ & 0.650 & 2.22 & 15.4 & \citet{lemon_2018} \\
\object{$\mathrm{J0140+4107}$} & $01^h 40^m 49^s.01$ & $+41^{\circ} 07' 59''.9$ & 2.500 & 1.44 & 11.6 & \citet{lemon_2018} \\
\object{$\mathrm{J0229+0320}$} & $02^h 29^m 58^s.21$ & $+03^{\circ} 20' 30''.10$ & 1.430 & 2.08 & 17.6 & \citet{lemon_2020} \\
\object{$\mathrm{J0252-3249}$} & $02^h 52^m 57^s.86$ & $-32^{\circ} 49' 08''.60$ & 2.240 & 2.10 & 17.3 & \citet{morgan_2000} \\
\object{$\mathrm{J0254-2443}$} & $02^h 54^m 17^s.28$ & $-22^{\circ} 43' 53''.52$ & 2.040 & 2.32 & 19.4 & \citet{lemon_2020} \\
\object{$\mathrm{J0302-0019}$} & $03^h 04^m 49^s.86$ & $-00^{\circ} 08' 13''.48$ & 3.286 & 2.90 & 21.7 & \citet{husemann_2018} \\
\object{$\mathrm{J0313-2546}$} & $03^h 13^m 38^s.12$ & $-25^{\circ} 46' 30''.49$ & 1.960 & 2.20 & 18.5 & \citet{lemon_2020} \\
\object{$\mathrm{J0322+0211}$} & $03^h 22^m 56^s.53$ & $+02^{\circ} 11' 58''.3$ & 3.510 & 1.10 & 8.0 & \citet{perna_2023} \\
\object{$\mathrm{J0332-2748}$} & $03^h 32^m 39^s.73$ & $-27^{\circ} 48' 50''.94$ & 3.060 & 0.70 & 5.4 & \citet{perna_2023} \\
\object{$\mathrm{J0402-3237}$} & $04^h 02^m 15^s.06$ & $-32^{\circ} 37' 33''.44$ & 1.280 & 0.65 & 5.4 & \citet{scialpi_2024} \\
\object{$\mathrm{J0443-2403}$} & $04^h 43^m 55^s.26$ & $-24^{\circ} 03' 26''.08$ & 1.780 & 1.85 & 15.6 & \citet{lemon_2020} \\
\object{$\mathrm{J0544-5922}$} & $05^h 44^m 30^s.63$ & $-59^{\circ} 22' 38''.70$ & 1.319 & 1.24 & 10.4 & \citet{aguita_2018} \\
\object{$\mathrm{J0740+2926}$} & $07^h 40^m 13^s.45$ & $+29^{\circ} 26' 48''.4$ & 0.980 & 2.60 & 20.7 & \citet{green_2011} \\
\object{$\mathrm{J0749+2255}$} & $07^h 49^m 22^s.97$ & $+22^{\circ} 55' 11''.80$ & 2.170 & 0.46 & 3.8 & \citet{chen_2023_vodka} \\
\object{$\mathrm{J0812+3349}$} & $08^h 12^m 54^s.83$ & $+33^{\circ} 49' 50''.23$ & 1.490 & 1.99 & 16.8 & \citet{lemon_2018} \\
\object{$\mathrm{J0818+0601}$} & $08^h 18^m 30^s.46$ & $+06^{\circ} 01' 38''.0$ & 2.361 & 1.10 & 9.0 & \citet{more_2016} \\
\object{$\mathrm{J0821+0735}$} & $08^h 21^m 43^s.36$ & $+07^{\circ} 35' 45''.9$ & 2.387 & 1.30 & 10.6 & \citet{more_2016} \\
\object{$\mathrm{J0840-0529}$} & $08^h 40^m 04^s.15$ & $-05^{\circ} 29' 11''.02$ & 1.386 & 0.38 & 3.2 & \citet{scialpi_2024} \\
\object{$\mathrm{J0841+4825}$} & $08^h 41^m 29^s.77$ & $+48^{\circ} 25' 48''.40$ & 2.950 & 0.46 & 3.6 & \citet{mannucci_2022} \\
\object{$\mathrm{J0847-0013}$} & $08^h 47^m 10^s.41$ & $-00^{\circ} 13' 02''.67$ & 0.626 & 1.00 & 6.8 & \citet{silverman_2020} \\
\object{$\mathrm{J0930+4614}$} & $09^h 30^m 21^s.16$ & $+46^{\circ} 14' 22''.8$ & 2.394 & 1.50 & 12.2 & \citet{more_2016} \\
\object{$\mathrm{J0932+0722}$} & $09^h 32^m 07^s.15$ & $+07^{\circ} 22' 51''.3$ & 1.994 & 1.42 & 11.9 & \citet{inada_2008} \\
\object{$\mathrm{J0942+2310}$} & $09^h 42^m 34^s.97$ & $+23^{\circ} 10' 31''.10$ & 1.833 & 2.45 & 20.7 & \citet{inada_2012} \\
\object{$\mathrm{J1000+0220}$} & $10^h 00^m 23^s.76$ & $+02^{\circ} 20' 37''.0$ & 7.150 & 0.26 & 1.3 & \citet{ubler_2024} \\
\object{$\mathrm{J1001+0136}$} & $10^h 01^m 05^s.18$ & $+01^{\circ} 36' 51''.26$ & 3.510 & 1.40 & 10.2 & \citet{perna_2023} \\
\object{$\mathrm{J1008+0351}$} & $10^h 08^m 59^s.55$ & $+03^{\circ} 51' 04''.4$ & 1.750 & 0.98 & 8.3 & \citet{inada_2008} \\
\object{$\mathrm{J1012+0335}$} & $10^h 12^m 54^s.74$ & $+03^{\circ} 35' 48''.93$ & 3.166 & 2.10 & 15.9 & \citet{herwig_2024} \\
\object{$\mathrm{J1034+0701}$} & $10^h 34^m 51^s.47$ & $+07^{\circ} 01' 21''.2$ & 1.250 & 3.10 & 25.9 & \citet{hennawi_2006} \\
\object{$\mathrm{J1035+0752}$} & $10^h 35^m 19^s.37$ & $+07^{\circ} 52' 58''.0$ & 1.220 & 2.70 & 22.4 & \citet{hennawi_2006} \\
\object{$\mathrm{J1043+4320}$} & $10^h 43^m 24^s.87$ & $+43^{\circ} 20' 49''.4$ & 2.225 & 1.70 & 14.0 & \citet{more_2016} \\
\object{$\mathrm{J1048+4541}$} & $10^h 48^m 20^s.91$ & $+45^{\circ} 41' 41''.25$ & 1.441 & 0.32 & 2.7 & \citet{mannucci_2023} \\
\object{$\mathrm{J1053+5001}$} & $10^h 53^m 20^s.15$ & $+50^{\circ} 01' 46''.02$ & 3.078 & 2.10 & 16.1 & \citet{hennawi_2010} \\
\object{$\mathrm{J1103+2348}$} & $11^h 03^m 08^s.12$ & $+23^{\circ} 48' 06''.19$ & 1.441 & 0.52 & 4.4 & \citet{mannucci_2023} \\
\object{$\mathrm{J1120+6711}$} & $11^h 20^m 12^s.11$ & $+67^{\circ} 11' 16''.0$ & 1.490 & 1.71 & 14.5 & \citet{pindor_2006} \\
\object{$\mathrm{J1124+5710}$} & $11^h 24^m 55^s.24$ & $+57^{\circ} 10' 56''.5$ & 2.310 & 2.20 & 18.0 & \citet{hennawi_2006} \\
\object{$\mathrm{J1138+6807}$} & $11^h 38^m 09^s.21$ & $+68^{\circ} 07' 38''.8$ & 0.770 & 2.60 & 19.3 & \citet{hennawi_2006} \\
\object{$\mathrm{J1139+4143}$} & $11^h 39^m 47^s.06$ & $+41^{\circ} 43' 51''.2$ & 2.230 & 2.30 & 19.0 & \citet{lemon_2018} \\
\object{$\mathrm{J1158+1235}$} & $11^h 58^m 22^s.99$ & $+12^{\circ} 35' 20''.31$ & 0.599 & 3.60 & 24.0 & \citet{green_2011} \\
\object{$\mathrm{J1158+1355}$} & $11^h 58^m 51^s.05$ & $+13^{\circ} 55' 35''.98$ & 2.062 & 3.24 & 27.0 & \citet{eftekharzadeh_2017} \\
\object{$\mathrm{J1202-0725}$} & $12^h 05^m 23^s.15$ & $-07^{\circ} 42' 33''.0$ & 4.700 & 3.50 & 22.7 & \citet{zamora_2024} \\
\object{$\mathrm{J1212+0912}$} & $12^h 12^m 44^s.33$ & $+09^{\circ} 12' 08''.10$ & 1.686 & 2.04 & 17.3 & \citet{inada_2008} \\
\object{$\mathrm{J1214+0102}$} & $12^h 14^m 05^s.12$ & $+01^{\circ} 02' 05''.18$ & 0.493 & 2.18 & 13.2 & \citet{silverman_2020} \\
\object{$\mathrm{J1215-0148}$} & $12^h 15^m 03^s.42$ & $-01^{\circ} 48' 59''.3$ & 6.050 & 2.00 & 11.4 & \citet{matsuoka_2024} \\
\object{$\mathrm{J1220+1126}$} & $12^h 20^m 16^s.9$ & $+11^{\circ} 26' 27''.09$ & 1.889 & 0.26 & 2.2 & \citet{glikman_2023} \\
\object{$\mathrm{J1235+0434}$} & $12^h 35^m 55^s.27$ & $+68^{\circ} 36' 27''.07$ & 1.529 & 3.51 & 29.7 & \citet{eftekharzadeh_2017} \\
\object{$\mathrm{J1236+5220}$} & $12^h 36^m 35^s.14$ & $+52^{\circ} 20' 58''.81$ & 2.570 & 3.07 & 24.6 & \citet{lusso_2018} \\
\object{$\mathrm{J1238+0105}$} & $12^h 38^m 21^s.66$ & $+01^{\circ} 05' 18''.6$ & 3.120 & 1.01 & 7.7 & \citet{tang_2021} \\
\object{$\mathrm{J1254+0846}$} & $12^h 54^m 55^s.12$ & $+08^{\circ} 46' 54''.18$ & 0.440 & 3.80 & 21.6 & \citet{green_2011} \\
\object{$\mathrm{J1259+1241}$} & $12^h 59^m 55^s.62$ & $+12^{\circ} 41' 53''.80$ & 2.190 & 3.55 & 29.4 & \citet{hennawi_2006} \\
\object{$\mathrm{J1309+5617}$} & $13^h 09^m 27^s.55$ & $+56^{\circ} 17' 38''.90$ & 2.515 & 2.80 & 22.6 & \citet{more_2016} \\
\object{$\mathrm{J1331+0335}$} & $13^h 31^m 45^s.98$ & $+03^{\circ} 35' 46''.20$ & 2.580 & 3.30 & 26.5 & \citet{lusso_2018} \\
\object{$\mathrm{J1337+6012}$} & $13^h 37^m 13^s.13$ & $+60^{\circ} 12' 06''.59$ & 1.721 & 3.12 & 26.4 & \citet{eftekharzadeh_2017} \\
\object{$\mathrm{J1339+1310}$} & $13^h 39^m 07^s.13$ & $+13^{\circ} 10' 39''.60$ & 2.238 & 2.10 & 17.3 & \citet{lusso_2018} \\
\object{$\mathrm{J1349+1227}$} & $13^h 49^m 29^s.84$ & $+12^{\circ} 27' 07''.00$ & 1.720 & 3.00 & 25.4 & \citet{hennawi_2006} \\
\object{$\mathrm{J1405+1350}$} & $14^h 05^m 56^s.92$ & $+13^{\circ} 50' 38''.3$ & 2.350 & 1.70 & 13.9 & \citet{more_2016} \\
\object{$\mathrm{J1416+0033}$} & $14^h 16^m 37^s.44$ & $+00^{\circ} 33' 52''.25$ & 0.433 & 0.66 & 3.7 & \citet{silverman_2020} \\
\object{$\mathrm{J1418-1610}$} & $14^h 18^m 17^s.68$ & $-16^{\circ} 10' 08''.29$ & 1.130 & 2.41 & 19.8 & \citet{lemon_2019} \\
\object{$\mathrm{J1426+3533}$} & $14^h 26^m 07^s.6$ & $+35^{\circ} 33' 51''.0$ & 1.175 & 0.69 & 5.7 & \citet{barrows_2012} \\
\object{$\mathrm{J1428+0500}$} & $14^h 28^m 55^s.44$ & $+05^{\circ} 00' 19''.89$ & 1.380 & 2.23 & 18.8 & \citet{lemon_2019} \\
\object{$\mathrm{J1433+1450}$} & $14^h 33^m 51^s.09$ & $+14^{\circ} 50' 05''.6$ & 1.506 & 3.34 & 28.3 & \citet{inada_2012} \\
\object{$\mathrm{J1444+5413}$} & $14^h 44^m 22^s.59$ & $+54^{\circ} 13' 20''.64$ & 1.584 & 3.45 & 29.2 & \citet{eftekharzadeh_2017} \\
\object{$\mathrm{J1502+1115}$} & $15^h 02^m 43^s.09$ & $+11^{\circ} 15' 57''.03$ & 0.390 & 1.40 & 7.4 & \citet{fu_2011} \\
\object{$\mathrm{J1508+3328}$} & $15^h 08^m 42^s.21$ & $+33^{\circ} 28' 02''.60$ & 0.880 & 2.90 & 22.4 & \citet{hennawi_2006} \\
\object{$\mathrm{J1540+4445}$} & $15^h 40^m 25^s.82$ & $+44^{\circ} 45' 16''.5$ & 0.610 & 2.74 & 18.5 & \citet{lemon_2018} \\
\object{$\mathrm{J1554+5817}$} & $15^h 54^m 18^s.49$ & $+58^{\circ} 17' 46''.89$ & 1.490 & 1.39 & 11.8 & \citet{lemon_2019} \\
\object{$\mathrm{J1600+0000}$} & $16^h 00^m 15^s.50$ & $+00^{\circ} 00' 45''.5$ & 1.010 & 1.90 & 15.3 & \citet{hennawi_2006} \\
\object{$\mathrm{J1606+2900}$} & $16^h 06^m 03^s.03$ & $+29^{\circ} 00' 51''.026$ & 0.770 & 3.50 & 25.9 & \citet{green_2011} \\
\object{$\mathrm{J1613+1708}$} & $16^h 13^m 20^s.01$ & $+17^{\circ} 08' 39''.40$ & 1.544 & 0.76 & 6.4 & \citet{ciurlo_2023} \\
\object{$\mathrm{J1643+3156}$} & $16^h 43^m 11^s.33$ & $+31^{\circ} 56' 19''.0$ & 0.586 & 2.30 & 15.2 & \citet{brotherton_1999} \\
\object{$\mathrm{J1821+6005}$} & $18^h 21^m 30^s.25$ & $+60^{\circ} 05' 26''.79$ & 2.050 & 1.48 & 12.4 & \citet{lemon_2018} \\
\object{$\mathrm{J2032-2358}$} & $20^h 32^m 37^s.78$ & $-23^{\circ} 58' 22''.5$ & 1.640 & 1.91 & 16.2 & \citet{lemon_2018} \\
\object{$\mathrm{J2037-4537}$} & $20^h 37^m 40^s.29$ & $-45^{\circ} 37' 57''.7$ & 5.660 & 1.24 & 7.3 & \citet{yue_2021} \\
\object{$\mathrm{J2057+0217}$} & $20^h 57^m 52^s.07$ & $+02^{\circ} 17' 48''.6$ & 1.520 & 1.06 & 9.0 & \citet{lemon_2018} \\
\object{$\mathrm{J2141-4629}$} & $21^h 41^m 48^s.85$ & $-46^{\circ} 29' 45''.7$ & 1.762 & 0.91 & 7.7 & \citet{aguita_2018} \\
\object{$\mathrm{J2209+0045}$} & $22^h 09^m 06^s.91$ & $+00^{\circ} 45' 43''.9$ & 0.445 & 1.67 & 9.6 & \citet{tang_2021} \\
\object{$\mathrm{J2215-3056}$} & $22^h 15^m 25^s.6$ & $-30^{\circ} 56' 34''.9$ & 1.347 & 0.71 & 6.0 & \citet{schechter_2017} \\
\object{$\mathrm{J2215-5204}$} & $22^h 15^m 40^s.10$ & $-52^{\circ} 04' 04''.44$ & 2.350 & 2.74 & 22.4 & \citet{lemon_2020} \\
\object{$\mathrm{J2228-3047}$} & $22^h 28^m 49^s.43$ & $-30^{\circ} 47' 34''.48$ & 1.949 & 0.38 & 3.2 & \citet{scialpi_2024} \\
\object{$\mathrm{J2250-6047}$} & $22^h 50^m 07^s.92$ & $-60^{\circ} 47' 23''.1$ & 1.080 & 2.02 & 16.5 & \citet{aguita_2018} \\
\object{$\mathrm{J2329-0522}$} & $23^h 29^m 15^s.07$ & $-05^{\circ} 22' 35''.88$ & 4.850 & 1.55 & 9.9 & \citet{yue_2023} \\
\object{$\mathrm{J2335+3201}$} & $23^h 35^m 22^s.52$ & $+32^{\circ} 01' 09''.08$ & 0.904 & 0.54 & 4.2 & \citet{ciurlo_2023} \\
\object{$\mathrm{J2336-0107}$} & $23^h 36^m 46^s.2$ & $-01^{\circ} 07' 32''.6$ & 1.285 & 1.67 & 14.0 & \citet{gregg_2002} \\
\object{$\mathrm{J2337+0056}$} & $23^h 37^m 13^s.66$ & $+00^{\circ} 56' 10''.8$ & 0.708 & 2.21 & 15.9 & \citet{tang_2021} \\

\end{longtable}

%

\end{appendix}

\end{document}